\documentclass[useAMS,usenatbib]{mn2e}
\usepackage{epsfig}
\usepackage{color}
\usepackage{natbib}
\usepackage{deluxetable}
\usepackage{amssymb}
\usepackage{amsmath}
\usepackage{graphicx}
\usepackage[colorlinks=true,citecolor=blue]{hyperref}
\def\lya{Lyman-$\alpha$ }
\def\lyb{Lyman-$\beta$ }

\def\Mpc{\hbox{Mpc}}

\def\cm2{cm$^{-2}$}

\def\lya{lyman-$\alpha$}

\def\Obh2{\Omega_{\rm b}h^{2}}

\def\J21{J_{21}}

\def\Mpc{Mpc}

\newcommand{\bit}{\begin{itemize}}
\newcommand{\eit}{\end{itemize}}
\newcommand{\ben}{\begin{enumerate}}
\newcommand{\een}{\end{enumerate}}
\def\lya{Ly$\alpha$ } 
\newcommand{\be}{\begin{equation}} 
\newcommand{\en}{\end{equation}}

\def\lya{Ly$\alpha$ } 
\def\lyb{Ly$\beta$ } 
\def\h2{H$_2$} 
\def\hi{H{\sc i}~}

\def\kms{kms$^{-1}$}

\twocolumn

\title[Ratio of median pixel optical depth profile]{Constraining the
 ratio of median pixel optical depth profile around $z\sim4$ quasars using the longitudinal proximity effect}

\author[Jalan, Chand \& Srianand] {{Priyanka
   Jalan$^{1,2}$\thanks{E-mail: priyajalan14@gmail.com }, Hum
   Chand$^{3,1}$, Raghunathan Srianand$^{4}$ } \\ $^{1}$Aryabhatta
  Research Institute of Observational Sciences (ARIES), Manora Peak,
  Nainital, 263002 India,\\ $^2$Department of Physics and
  Astrophysics, University of Delhi, Delhi 110007,
  India\\ $^3$Department of Physics and Astronomical Sciences, Central
  University of Himachal Pradesh (CUHP), Dharamshala-176215, India
  \\ $^{4}$Inter-University Centre for Astronomy and Astrophysics
  (IUCAA), Postbag 4, Ganeshkhind, Pune 411 007, India}
\begin{document}
\date{Accepted ---. Received ---; in original form ---}

\pagerange{\pageref{firstpage}--\pageref{lastpage}} \pubyear{2020}

\maketitle

\label{firstpage}
\begin{abstract}
We present a detailed study of the longitudinal proximity effect using
a sample of 85 quasars spanning an emission redshift range of $3.5
\leq z_{em} \leq 4.5$ and Lyman continuum luminosity ($L_{912}$)
ranging from 1.06$\times 10^{31}$ to 2.24$\times 10^{32}$ erg s$^{-1}$
Hz$^{-1}$. We use the high-quality spectra of these quasars obtained
at a spectral resolution of $R\sim$ 5100 and S/N $\sim$ 30 using
X-SHOOTER spectrograph mounted on the Very Large Telescope (VLT). In
our analysis, we compared the transmitted flux and pixel optical
  depth of the \lya absorption originating from the vicinity of
quasars to those from the general intergalactic medium by using a
redshift matched control sample. The longitudinal proximity effect is found up to $r \leq 12$ Mpc (proper) from
quasars. By appropriately scaling up the  pixel optical depth in
the vicinity of quasars to account for the excess ionization by
quasars, we constrain the ratio of median \hi optical depth in the
vicinity of the quasar to that of the IGM ($R_\tau (r)$). The $R_\tau (r)$ is found to be significantly
higher than unity up to 6 Mpc from the quasar with a typical radial
profile of the form $R_\tau(r) = 1+A \times exp(-r/r_0)$ with
  $A=9.16\pm 0.68$ and $r_0= 1.27\pm 0.08$ Mpc. The integrated value
of the scaled pixel optical depth over the radial bin of 0-6 Mpc
is found to be higher by a factor of $2.55\pm 0.17$ than the
corresponding integrated value of the median pixel optical depth
of the IGM. We also found $R_\tau (r)$ to be luminosity dependent.
\end{abstract}


\section{Introduction}
\label{s:intro}
Previous studies using spectra of distant quasars have established that the number of \lya forest absorption lines per unit redshift generally increases with increasing redshift \citep[e.g.,   see][]{Lu1991ApJ...367...19L,Giallongo1991MNRAS.251..541G,Kulkarni1993ApJ...413L..63K,Bechtold1994ApJS...91....1B,Giallongo1996ApJ...466...46G,Scott2000ApJS..130...67S,Kim2007MNRAS.382.1657K,Agliolpe32008AA...480..359D,Calverley2011MNRAS.412.2543C,Partl2011MNRAS.415.3851P,Becker2013MNRAS.430.2067B,Becker2015PASA...32...45B,Monzon2020AJ....160...37M}. However, in the vicinity of a quasar (up to several proper Mpc), the ionizing photons from the quasar will dominate the ionization state of the gas over those from the ultraviolet background (UVB) radiation. This produces a region of less \lya absorption (i.e., enhanced transmitted flux) near the emission redshift ($z_{em}$) of the quasar, first noted by \citet[][]{Carswell1982MNRAS.198...91C}, known as the {\it proximity effect} \citep[e.g., see][]{Murdoch1986ApJ...309...19M,Tytler1987ApJ...321...69T,Bajtlik1988ApJ...327..570B,Kulkarni1993ApJ...413L..63K,Bechtold1994ApJS...91....1B,srianand1996MNRAS.280..767S,Cooke1997MNRAS.284..552C,Liske2001MNRAS.328..653L,Worseck2006A&A...450..495W,Faucherlpe22008ApJ...673...39F,Wildlpe12008MNRAS.388..227W,Prochaska2013ApJ...776..136P,Khrykin2016ApJ...824..133K,Jalan2019ApJ...884..151J}. 
\par The distance up to which the effect of quasar's radiation dominates, depends on the intensity of UVB and \hi ionizing luminosity of the quasar. Therefore, the proximity effect has been used by many previous studies to get an estimate of the strength of the UVB radiation at various redshifts \citep[e.g., see][]{Bajtlik1988ApJ...327..570B,Kulkarni1993ApJ...413L..63K,Giallongo1996ApJ...466...46G,Cooke1997MNRAS.284..552C,Scott2000ApJS..130...67S,Liske2001MNRAS.328..653L,Agliolpe32008AA...480..359D}. However, most of these classical proximity effect analysis assume that the density distribution of gas in the vicinity of a quasar is similar to that far away from it i.e., the general intergalactic medium \citep[IGM; for more details see also,][]{Faucherlpe22008ApJ...673...39F,Calverley2011MNRAS.412.2543C,Partl2011MNRAS.415.3851P}. This
could cause the intensity of the UVB radiation to be overestimated by up to a factor of 2-3 when measured using the proximity effect
\citep[][]{Loeb1995ApJ...448...17L,Faucherlpe22008ApJ...673...39F}. 

\par However, instead of measuring the UVB intensity from the proximity effect analysis, the procedure can be reversed to estimate the density profile around the quasar \citep[e.g., see][]{Scott2000ApJS..130...67S,Rollinde2005MNRAS.361.1015R,Guimaraes2007MNRAS.377..657G,Dodorico2008MNRAS.389.1727D}, if the UVB measurements are known from an independent method \citep[][and references therein]{Faucher2009ApJ...703.1416F,Haardt2012ApJ...746..125H,Khaire2015MNRAS.451L..30K,Khaire2019MNRAS.484.4174K}. Additionally, the gas distribution in the transverse direction of the quasar can be studied using multiple sightlines with small separations. The expected decrement of the \lya absorption lines in the spectrum of the background quasar near the emission redshift of the foreground quasar induced by the ionizing radiation from the foreground quasar in its transverse direction is known as transverse proximity effect \citep[e.g., see][]{srianand1997ApJ...478..511S,Adelberger2004ApJ...612..706A,Schirber2004ApJ...610..105S,Rollinde2005MNRAS.361.1015R,Worseck2007A&A...473..805W,Goncalves2008ApJ...676..816G,Gallerani2011JPhCS.280a2008G,Hennawi2013ApJ...766...58H,Schmidt2018ApJ...861..122S,Jalan2019ApJ...884..151J}.

\par In our previous work, \citet[][hereafter, JCS19]{Jalan2019ApJ...884..151J}, we studied the proximity effect in the longitudinal and transverse direction for quasars at $2.5\leq z \leq 3.5$. We have used 181 quasar pairs with separation $<$ 1.5 arcmin from SDSS-DR12 \citep[for more information, see][]{Paris2017A&A...597A..79P} quasar catalogue. Our analysis includes a novel technique of studying the proximity effect using a control sample of quasars, matching in the absorption redshift and continuum signal-to-noise ratio (S/N), to take into account the effect of optical depth evolution with the redshift. We compared the transmitted flux and/or optical depth from \lya absorption lines originating in the vicinity and far from the quasar i.e., IGM. We detected enhancement and decrement of the transmitted flux within a radial distance of 4 Mpc from the quasar in the longitudinal and transverse direction respectively. Additionally, JCS19 has also taken into account the effect of spectral resolution and optical depth while lifting the degeneracy between excess ionization from the quasars and the excess density of the absorbing gas based on their detailed simulations. The ratio of median pixel optical depth in the longitudinal direction after applying the ionization correction is found to be consistent with that in the transverse direction without applying any correction for ionization by the quasars. We interpreted this as an indication for the presence of an anisotropic obscuration in the transverse direction (e.g., by dusty torus) with $\leq$27\% (at $3 \sigma$ confidence level) quasar's ionization/illumination as compared to its longitudinal direction. We found that our sample was dominated by Type-I AGNs which supported the results \citep[also see,][]{Prochaska2014ApJ...796..140P, Lau2016ApJS..226...25L,Lau2018ApJ...857..126L}. 

\par The main focus of our previous work (JCS19) was to carry out the comparison of the gas distribution around quasars at $\left<z_{em} \right> \sim 3$ along the longitudinal and transverse directions. In this paper, we aim to extend our analysis of the longitudinal proximity effect to higher redshifts by applying a similar technique as used in JCS19. However, due to the steep rise in the number of H~{\sc i} clouds with redshifts the study of longitudinal proximity effect at higher redshift can easily be affected by line blending at a typical spectral resolution of large surveys such as SDSS ($R\sim$ 2000). Recently, \citet[][]{Lopez_2016refId0} have released a sample of 100 quasars (hereafter, XQ-100 survey) observed with X-SHOOTER spectrograph \citep[e.g., see][]{Vernet2011A&A...536A.105V} mounted on Very Large Telescope (VLT). Quasars in this sample have $z_{em}$ from 3.5 to 4.71, and the available spectra are of high spectral resolution ($R\sim$ 5000-9000) and high S/N (median $\sim$ 30). The advantage of this sample is that due to the large spectral coverage of the X-SHOOTER spectrograph, it is possible to observe Mg~{\sc ii}, [O~{\sc iii}] and [O~{\sc ii}] emission lines that are redshifted to the NIR region for quasars at these redshifts. These emission lines can be used to compute the systemic redshift of the quasars. Thus, the systemic redshifts can be estimated with higher accuracy corresponding to velocity dispersion of $<$ 60 \kms~($\Delta v = c \times \Delta z/[1+z]$). This accuracy is much higher in comparison to the accuracy achieved using the emission lines lying in the optical region for these quasars. As pointed out by \citet[][]{Shen2016ApJ...831....7S} redshift measured using C~{\sc iii}], He~{\sc ii}, C~{\sc iv} (after correcting for the luminosity-dependent blueshift) and Mg~{\sc ii} lines have the systemic velocity shift of $\sim$ 230 \kms, $-$167 \kms, $-$27 \kms, $-$57 \kms~and scatter of $\sim$ 233 \kms, 242 \kms, 415 \kms~and 205 \kms, respectively \citep[see also,][]{Denney2016ApJS..224...14D,Eilers2017ApJ...840...24E,Dix2020ApJ...893...14D}. Therefore, the availability of the high-quality spectra provided by XQ-100 survey at $z_{em} > 3.5$ in conjunction to the aforementioned new technique devised by JCS19 motivates our analysis to estimate the radial profile of the median pixel optical depth around these high-z quasars in comparison to the median  pixel optical depth of the general IGM.
\par The paper is structured as follows. In Sect.~\ref{s:sample}, we discuss our sample and its properties. In Sect.~\ref{s:Analysis}, we present transmitted flux analysis and results. This is followed by estimating an appropriate ionization correction using simulated spectra for the pixel optical depth values when observed using a moderate resolution spectrograph. This relation is then applied to the real data to constrain the median pixel optical depth radial profile around high-z quasars. The discussions and conclusions are presented in Sect.~\ref{s:Discussion}. Throughout this paper, we have used a flat background cosmology with cosmological parameters $\Omega_m$, $\Omega_\lambda$ and $H_o$ to be 0.286, 0.714, 69.6 \kms \Mpc$^{-1}$, respectively, \citep[][]{Bennett2014ApJ...794..135B}. Moreover, all the distances mentioned in this paper are proper distances unless noted otherwise.

\section{Data and its properties}
\label{s:sample}
\begin{figure}	
  \centering
  \includegraphics[height=7cm,width=8.0cm]{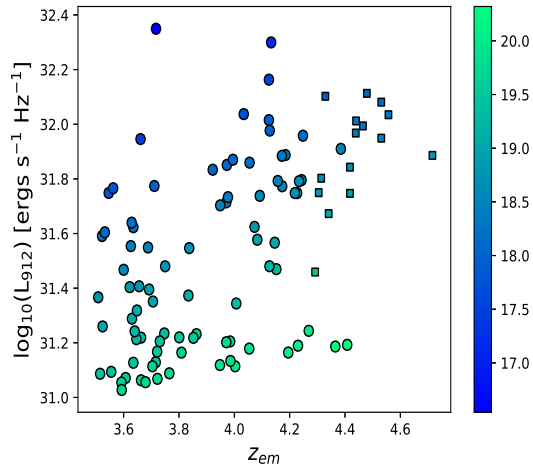}
  \caption {The plot shows the Lyman continuum luminosity ($L_{912}$)
   versus emission redshift ($z_{em}$) for the 100 quasars. The color-shade in the right side provides the V-band magnitude
   scale. The circle represents the 85 quasars used in our main
   sample (see Sect.~\ref{s:sample}). The square represents 15
   quasars which got excluded from our proximity analysis sample due
   to lack of control sample but they are used for the control sample (as
   elaborated in Sect.~\ref{s:control_sample}).}
  \label{Fig:z_lum}
\end{figure}

We use 100 quasar spectra from the XQ-100 survey, obtained using VLT/X-SHOOTER in the period between 2012-02-10 and 2014-02-23 within the ESO large program entitled ``Quasars and their absorption lines: a legacy survey of the high redshift universe with X-SHOOTER'' lead by \citet[][]{Lopez_2016refId0}. X-SHOOTER is a triple arm spectrograph, with UV-Blue\footnote{ To avoid confusion with UVB as used for ultraviolet background, the UV-blue arm is represented as UV-Blue.}, VIS (visible) and NIR (near-infrared) arm covering a spectral range of 315-560 nm, 540-1020 nm and 1000-2480 nm \citep[e.g., see also,][]{Vernet2011A&A...536A.105V}, respectively. Therefore, in one integration X-SHOOTER provides a spectral coverage from the $\sim$310 nm (atmospheric cutoff) to the 2480 nm (NIR) at a moderate resolution of R $\approx$ 5000-9000.  The coverage up to 2480 nm is available only for 47 of the quasars in the sample due to the employed K band blocking filter.  However, this spectral coverage leads to the full coverage of the \lya forest in the optical spectral regime for the quasars at higher redshifts and the [O~{\sc iii}], [O~{\sc ii}] and Mg~{\sc ii} emission lines in the NIR regime allows one to have a precise estimate of systemic redshifts of the quasars (e.g., see Sect.~\ref{s:zq}). The distribution of emission redshifts of these 100 quasars spanning a range of 3.51 to 4.71 is shown in Fig.~\ref{Fig:z_lum}. These are measured using principal component analysis (PCA). The details of the sample are fully described in the \citet[][]{Lopez_2016refId0}. \par For the analysis presented here, we mostly utilize the spectra taken in VIS and UV-Blue arms only. The resolving power in the UV-Blue and VIS arm is 5100 (pixel width of 20 \kms) and 8800 (pixel width of 11 \kms) respectively obtained from the header of the fits file and data release which is also consistent with \citet[][]{Walther2018ApJ...852...22W} findings. In order to have a similar resolution throughout the spectra, we have convolved the higher resolution spectra of VIS arm to a lower resolution value corresponding to the UV-Blue arm by using a Gaussian kernel with $FWHM = [FWHM_{UVB}^2-FWHM_{VIS}^2]^{1/2}$. We then interpolated flux, continuum, and error to a common wavelength grid of 20 \kms~per pixel using the ``cubic-spline interpolation method''.
 
 \par Therefore, the full spectral coverage of the \lya forest along with the high S/N (median S/N = 30) and high resolution (R $\sim$ 5100) achieved, clearly makes XQ-100 a unique data set to study the quasars proximity effect at high-z. Out of these 100 quasars, we could make use of 85 quasars as the main sample that are marked as flag=1 in Table~\ref{table:sources}. The remaining 15 quasars (marked as flag=0) were excluded from the proximity sample due to lack of IGM control sample for them but included in the control sample as discussed in Sect.~\ref{s:control_sample} and marked by squares in Fig.~\ref{Fig:z_lum}.

\begin{table*}
  \centering
    {
      \setlength{\tabcolsep}{1.0pt}
\caption{{Some properties of 100 quasars used in our analysis with masked absorption redshifts.}
\label{tab:source_info}}
\begin{tabular}{rrrrrrrcc}
\hline
 \multicolumn{1}{c}{SN.}  & \multicolumn{1}{c}{Quasar name}  &
 \multicolumn{1}{c}{$z_{em}$}  & \multicolumn{1}{c}{Ra} &
 \multicolumn{1}{c}{Dec} &  \multicolumn{1}{c}{$m_V$} &
 \multicolumn{1}{c}{log[$L_{912}]$}  & \multicolumn{1}{c}{Flag for main sample} &
 \multicolumn{1}{c}{$z_{a}$ (class)}
 \\
\multicolumn{1}{c}{}  & \multicolumn{1}{c}{}  & \multicolumn{1}{c}{}  &
\multicolumn{1}{c}{(deg)}  & \multicolumn{1}{c}{(deg)} & \multicolumn{1}{c}{} &
 \multicolumn{1}{c}{(erg~s$^{-1}$ Hz$^{-1}$)} & \multicolumn{1}{c}{=1 for 85 quasars} & \multicolumn{1}{c}{Masked absorbers in the \lya forest}
 \\
\multicolumn{1}{c}{[1]}  & \multicolumn{1}{c}{[2]}  & \multicolumn{1}{c}{[3]}  &
\multicolumn{1}{c}{[4]}  & \multicolumn{1}{c}{[5]} & \multicolumn{1}{c}{[6]} &
 \multicolumn{1}{c}{[7]} & \multicolumn{1}{c}{[8]} & \multicolumn{1}{c}{[9]}
 \\
 \\
 \hline
  1  &  J0003-2603  &  4.125   &   0.845    &  $-$26.055   &  17.530   &  32.163  &   1  &  3.39 (DLA), 3.39 (NAL SiIV])	\\										
  2  &  J0006-6208  &  4.440   &   1.715    &  $-$62.134   &  18.290   &  32.012  &   0  &  3.78 (DLA) \\
  3  &  J0030-5129  &  4.173   &   7.644    &  $-$51.495   &  18.570   &  31.773  &   1  &    --		\\	  	
  4  &  J0042-1020  &  3.863   &  10.582    &  $-$10.337   &  19.530   &  31.232  &   1  &  --		\\	   	
  5  &  J0048-2442  &  4.083   &  12.143    &  $-$24.702   &  18.940   &  31.577  &   1  &  3.76 (sub-DLA)	\\
  .. &  .....    & .....     &    .....   &  .....    &  .....   &  .....  &  ..... & ... ...  \\
   \\
  \hline
  
\multicolumn{9}{l}{{ Note.} Here, we show only a small portion of the
  table to display its form and contents. The entire table with
  details}\\ \multicolumn{9}{l}{ of all 100 quasars are available in an
  online version.}\\

\end{tabular}
\label{table:sources}
    }
\end{table*}

\subsection{Distances and luminosities}
\label{s:Properties}
 The Lyman continuum luminosity can be calculated using flux-calibrated spectra as detailed in JCS19. However, as mentioned by \citet[][]{Lopez_2016refId0} that the flux values in XQ-100 can be uncertain by an order-of-magnitude. Therefore, the Lyman continuum luminosity was calculated using the V-band magnitude of these quasars based on the formalism provided by \citet[][]{Liske2001MNRAS.328..653L}. Here, we assume the continuum to be a power-law i.e., $f_\nu \propto \nu^{-\alpha}$ with the flux density $f_\nu$ at the observed wavelength $\lambda$ as,
\begin{equation*}
f_\nu(\lambda) = \left[\frac{\lambda_V}{\lambda (1+z_{em})^{-1}}
  \right]^{-\alpha} (1+z_{em}) \times 10^{-0.4(m_V-k_V)} f_{\nu V}(0)
\end{equation*}
where $\lambda_V$, $k_V$, $m_V$ and $f_{\nu V}(0)$ are the central wavelength, $K$-correction \citep[e.g., see][]{Cristiani1990A&A...227..385C}, observed magnitude and 0-magnitude flux in the $V$-band respectively. Here, we have assumed
\begin{equation}
  \alpha \approx \biggl\{\begin{array}{ll}0.44~~;~\lambda > 1300~\AA
      \\  1.57~~;~\lambda \leq 1300~\AA \end{array}
  \label{eq:power}
\end{equation}
 \citep[e.g., see][and references
  therein]{Vanden2001AJ....122..549V,Tefler2002ApJ...565..773T,khaire2015ApJ...805...33K}. We note here that the UV-spectral index may vary from quasar to quasar and the impact of ignoring such dispersion was studied in our previous work (JCS19). We found that the effect of this variation is negligible as far as the recovered ratio of median pixel optical depth profile is concerned. Therefore, we have used the value of $\alpha$ as given in Eq.~\ref{eq:power}. 
  \par We also correct the observed flux value in our spectrum for Galactic extinction \citep[e.g., see][]{Schlegel1998ApJ...500..525S} by applying a correction factor
 $10^{0.4 A(\lambda)}$ where,
\begin{equation}
  A(\lambda) = R_V E(B-V) \frac{A(\lambda)}{A(V)}
\end{equation}
 using, $R_V= 3.1$ \citep[e.g., see][and references therein]{Clayton1988AJ.....96..695C,Liske2001MNRAS.328..653L}. The flux is calculated at the threshold wavelength of the H~{\sc i} ionizing photons (i.e., 912 \AA) and the corresponding luminosity as, $4\pi d_L^2 \times F_\nu(912$ \AA), where $d_L$ is the luminosity distance to the quasar. As a consistency check of the estimated Lyman continuum luminosity ($L_{912}$) based on this method, we also used the spectroscopic measurements to calculate it for the 53 sources for which SDSS flux-calibrated spectra are also available. Here, we used the flux at 1325 \AA~(in the rest frame of quasars) in conjunction with the spectral slope given by Eq.~\ref{eq:power} to compute the $F_\nu(912$ \AA) as detailed in JCS19. The same spectra are also used to estimate the V-band magnitude using the transfer function as given by \citet[][]{Johnson1951ApJ...114..522J}. The Lyman continuum luminosity estimated using the spectrum and V-band magnitude is found to be consistent with each other within 10\%. Therefore, for the sake of homogeneity, we have adopted the $L_{912}$ estimate based on the photometric method for all the sources in our sample. Here, a small caveat could be the difference in the epochs of photometric and spectroscopic observations over which the quasar's magnitude might have varied.  For instance, the photometric light curve available in Catalina Real-time Transient Survey \citep[CRTS\footnote{http://crts.caltech.edu},][]{Drake2009ApJ...696..870D} for 96 sources in our sample, observed over a period of about 8 years, shows a typical variation of $\sim$10\% in their magnitude. This could at the maximum lead to additional 10\% uncertainty in the estimated Lyman continuum luminosity.  \par In Fig.~\ref{Fig:z_lum}, we show $z_{em}$ versus $L_{912}$ for the quasars in our sample along with their V-band magnitude \citep[taken from the summary file provided by][]{Lopez_2016refId0}. It can be seen from the figure that for our sample the $L_{912}$ ranges from 1.06$\times 10^{31}$ to 2.24$\times 10^{32}$ erg s$^{-1}$ Hz$^{-1}$ and $z_{em}$ in the range from 3.51 to 4.71 (e.g., see Table~\ref{table:sources}). 
 \par The proper radial distance ($r$) between the quasar with an emission redshift of $z_{em}$ and an absorbing cloud at an absorption redshift
 of $z_{a}$, is computed as,
 \begin{equation}
        r(z_{em},z_{a}) =
        \frac{c \times [z_{a}-z_{em}]}{(1+z_{em})~H(z_{em})}=\frac{\Delta~v}{H(z_{em})}
        \label{eq:r_para}
      \end{equation}
 where, $H(z_{em}) = H_0\sqrt{\Omega_m(1+z_{em})^3+\Omega_\lambda}$ is the Hubble constant at $z_{em}$ \citep[][]{Kirkman2008MNRAS.391.1457K}. For  further analysis of longitudinal proximity effect, we consider the \lya forest within  $-50$ Mpc $\leq r(z_{em},z_{a}) \leq 0$ \Mpc~along the line of sight to the quasar as the proximity region. The \lya absorption seen  between the \lyb and \lya emission lines after excluding the proximity region is considered as the IGM in our study. The negative values of the distances given by Eq.~\ref{eq:r_para} for $z_{a}<z_{em}$ are just to indicate that absorbing clouds are towards the observers. However, in our calculations, we have used the absolute values for distances.

\begin{figure}	
  \includegraphics[height=7cm,width=8.0cm]{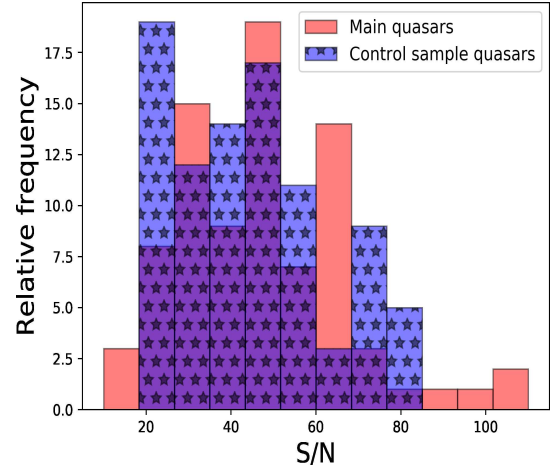}
  \includegraphics[height=7cm,width=8.0cm]{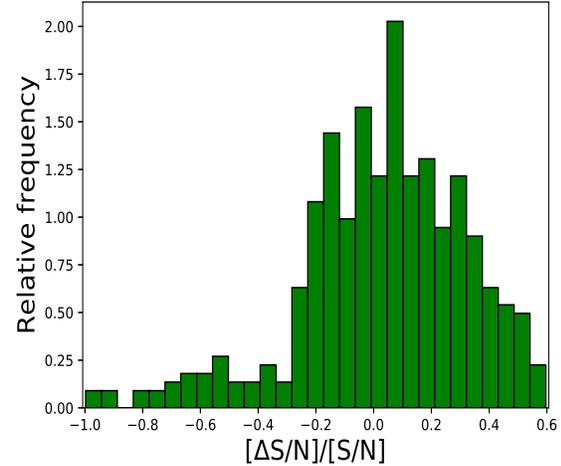}
  \caption {{\it Top panel:} The histogram plot showing the
   distribution of the S/N ($\langle$continuum/error$\rangle$) in the
   proximity region of the main quasars (S/N$_{p}$) and in the
   corresponding redshift region in the control sample
   (S/N$_{\textsc{igm}}$). The control sample histogram is normalized to that of
   the peak value of the main quasar's histogram. {\it Bottom panel:} The
   histogram plot showing the distribution of [$\Delta$S/N]/[S/N]
   ($\equiv$ [S/N$_{p}$-S/N$_{\textsc{igm}}$]/[S/N$_{p}]$).}
  \label{Fig:z_snr_match}
\end{figure}

\begin{figure*}	
  \centering \includegraphics[height=12cm,width=15.0cm]{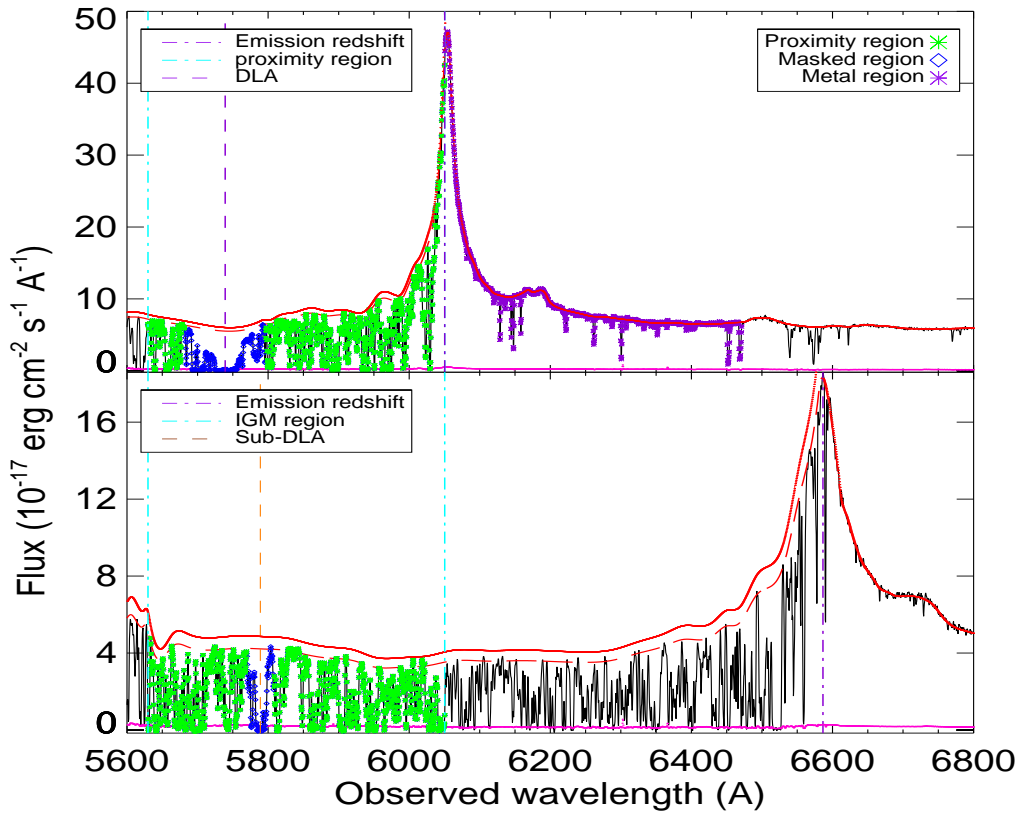}
  \caption {{\it Upper panel:} Spectrum of one of the quasars
    (J021429.4-051745.4) at $z_{em}$ = 3.977 (purple, dot-dashed
    vertical line shows the \lya emission) in our sample. The dashed and solid red curve shows the un-absorbed continuum flux before and after applying the continuum correction respectively (e.g., see Sect.~\ref{s:conti}). The vertical
    dashed lines (cyan) represent the 50 Mpc proximity region. The
    noise spectrum is also plotted (solid magenta). After masking the
    $\pm 3000$ \kms~region around the absorption redshift of the DLA
    (purple-dashed line), the pixels considered in the proximity
    region are shown in dots (green). The purple asterisk sign shows
    50 Mpc region redward of the \lya emission line. {\it Lower
      panel:} The plot shows the spectrum of one of the quasars
    (J052915.9-352601.2) in the corresponding control sample of the
    above main quasar (e.g., see Sect.~\ref{s:control_sample}). The
    purple dot-dashed line shows the \lya emission while the
    cyan-dashed line shows its IGM region corresponding to the above
    proximity region. After masking the $\pm 1000$ \kms~region around
    the absorption redshift of the sub-DLA (brown-dashed line), the
    pixels considered in the IGM region are shown in dots (green).}
  \label{Fig:spectra}
\end{figure*}

 \subsection{Control sample of \lya forest}
 \label{s:control_sample}

 In this paper, we carry out a statistical analysis by using the transmitted flux $(F_{t}[\lambda_i])$ and pixel optical depth,  $\tau(\lambda_i)$ of the \lya absorption (e.g., see JCS19 and references therein), i.e.,
 \begin{equation}
  F_{t}(\lambda_i) \equiv
  F(\lambda_i)/F_{c}(\lambda_i) = e^{-\tau(\lambda_i)}
  \label{eq:ft}
 \end{equation}
 where, $\tau(\lambda_i)$ is the pixel optical depth-integrated over the pixel width of the observed spectrum (i.e., 20 \kms). Here,
 $F(\lambda_i)$ and $F_{c}(\lambda_i)$ are the observed flux and the unabsorbed continuum flux respectively at the $i^{th}$ pixel having a wavelength $\lambda_i$. In our analysis, the pixels with $F_{t}(\lambda_i) \geq 1$ and $F_{t}(\lambda_i) \leq 0$ are also included. However, since the optical depth values for $F_{t}(\lambda_i) \leq 0$ are ill-defined, we assign $F_{t}(\lambda_i) = \sigma(\lambda_i)$ for such pixels, where $\sigma(\lambda_i)$ represents the corresponding flux measurement error. Moreover, the percentage of such pixels are very small ($\sim 1\%$), therefore, the inclusion and/or exclusion of these pixels will have a negligible effect on our further analysis. We statistically compare the distribution of the pixel optical depth and/or transmitted flux among the proximity region with that originating from the general IGM. However, in order to combine these values from various absorption redshifts one has to account for the strong redshift evolution of the optical depth. The evolution curve in principle allows one to scale the optical depth at various absorption redshifts in the proximity region to a reference redshift, and then carry out the statistical comparison of it with that of the average optical depth in the IGM at that reference redshift \citep[e.g., see][]{Rollinde2005MNRAS.361.1015R,Kirkman2008MNRAS.391.1457K}. However, this method does have a caveat that it may introduce  optical depth values by scaling up the continuum noises.
  
\par Therefore, instead of using this optical depth evolution curve to scale the optical depths at various redshifts to a reference redshift, we adopt an approach similar to JCS19 of using a control sample. In this regard, we construct a control sample, for each proximity region of a spectrum, consisting of those quasars in which their IGM (i.e., $r(z_{em},z_{a}) < -50$ Mpc) covers the same \lya absorption redshift region as that of the proximity region. The method has an advantage as it allows direct comparison of the observed optical depths in the proximity region and IGM region without using any scaling relationships. For constructing such a control sample, we made use of the \lya forest more than 50 Mpc away from our parent sample of 100 quasars. Out of the 100 main sightlines, we could not find a control sample for 7 sightlines with the aforementioned exact match in the redshift. This reduces our main sample to 93 sightlines. 
 
 \par In our search for the control sample, we noticed that for all quasars in our sample we could get typically about 5 sightlines of IGM. However, for a few of these sources, we could even get $>5$ sightlines in the control sample. For such cases, we picked up 5 sightlines having the closest match in S/N with the \lya forest in the proximity region. This also helps to ensure a good match in S/N as shown in the top panel of Fig.~\ref{Fig:z_snr_match}. Also, in the bottom panel of Fig.~\ref{Fig:z_snr_match}, we show the distribution of relative change of S/N of the proximity region (S/N$_p$) and IGM region (S/N$_{\textsc{igm}}$\footnote{ We assign ``p'' and ``igm'' as sub-scripts or super-scripts to represent the proximity sample and IGM sample respectively, throughout this paper.}) i.e., [$\Delta$S/N]/[S/N] ($\equiv$ [S/N$_{p}$-S/N$_{\textsc{igm}}$]/[S/N$_{p}]$) showing a $\leq$25\% deviation for about 68\% of our sample. For 8 quasars, we could not find a control sample with $\lvert$ [$\Delta$S/N]/[S/N] $\rvert <$ 1 and hence were excluded from our proximity analysis sample. This reduces our main sample from 93 sightlines to 85 sightlines. These 85 quasars are marked with flag=1 in column 8 of Table~\ref{tab:source_info}.
 
 \par In the lower panel of Fig.~\ref{Fig:spectra}, we show for illustration a spectrum used as a control sample, and the same for the main sample is shown in the upper panel. It shows the absorption redshift match between the two samples (green dots) with IGM being 50 Mpc away from the quasar's emission.  Additionally, all the samples used in our analysis were observed with the same spectral setting in X-SHOOTER. As a result, any effect of spectral resolution will either have a similar or no effect in both of the samples and hence will have minimal impact (if any) for our further analysis.

\subsection{DLA, sub-DLA and NAL}
The proximity effect analysis depends on the H~{\sc i} content in the neighbourhood of the quasar as compared to that of the intergalactic medium (IGM). Therefore, we mask the spectral region consisting of systems such as damped \lya (DLA)/sub-DLA, narrow-absorption lines (NAL) and Lyman limit systems (LLS) simply to avoid bias due to any strong absorption in our analysis.  \par To mask such absorption region, we have used \citet[][]{10.1093/mnras/stv2732} compilation, where they have provided a sample of 24 intervening DLA systems and 1 proximate DLA systems with column density (in units of cm$^{-2}$) log[N$_{\textsc{hi}}]> 20.3 $ towards these 85 quasars. Similarly, \citet[][]{10.1093/mnras/stz2012} provides a list of 207 H~{\sc i} absorbers with log[N$_{\textsc{hi}}]> 18.8 $ along these 85 quasars. Out of these 207 systems, 90 are LLS with 18.8 $\leq$ log[N$_{\textsc{hi}}] \leq 19 $, 88 are sub-DLAs with 19 $<$ log[N$_{\textsc{hi}}] \leq 20.3 $ and 22 are identified as DLAs \citep[already included in the][]{10.1093/mnras/stv2732}. In these catalogues, they have provided details such as absorption redshifts ($z_a$) and column densities of individual systems. We have used the $z_a$ information to mask a spectral region of $\pm$ 3000 \kms~and $\pm 1000$ \kms~around the absorption redshift of the DLAs and LLS/sub-DLAs, respectively. Additionally, \citet[][]{Perrotta2016MNRAS.462.3285P} listed the absorption redshift of the 833 NALs consisting of metal absorptions along the 85 quasars in our sample. The \lya absorption corresponding to these absorbers might be optically thin and hence can respond to the quasar's radiation \citep[e.g., see][]{Kim2016MNRAS.457.2005K}. Therefore, even though we do not mask the \lya region, however, if the corresponding metal lines such as C~{\sc iv}, Si~{\sc iv}, N~{\sc v}, C~{\sc ii} lies in the spectral region used in our analysis, we do mask a region of $\pm 500$ \kms~around these metal absorption lines. In Fig.~\ref{Fig:spectra}, we have marked for illustration the masked region in the proximity region of the main sample (top panel) and IGM region of the control sample (bottom panel).

 \section{Analysis}
 \label{s:Analysis}
 \subsection{Quasar Continuum}
 \label{s:conti}

 As evident from Eq.~\ref{eq:ft} that the analysis of the transmitted flux and the pixel optical depth can be significantly affected by any uncertainty in the continuum fit to the spectral region of the \lya forest. For the sample of XQ-100 quasar, \citet[][]{Lopez_2016refId0} have given the continuum fit based on the cubic spline method as shown with a red-dashed line in Fig.~\ref{Fig:spectra}. However, as pointed out by \citet[][]{Lidz2006ApJ...638...27L}, that such an algorithm has its caveat of underestimating continuum flux in the \lya forest at low-resolution with low S/N data and a careful visual check is required for each spectrum. Additionally, the continuum flux in the \lya forest at these redshifts, in general, might be underestimated due to the blending of multiple low optical depth \lya absorption lines \citep[e.g., see  also,][]{Seljak2003MNRAS.342L..79S,Lidz2006ApJ...638...27L,Faucherlpe22008ApJ...673...39F}. Therefore, we carried out a visual check of their continuum in the regions that are free from absorption and found that the r.m.s. of the fit is better than the 1/[S/N] allowed variation. We note here that such an underestimation of continuum flux will lead to underestimation of the optical depth both in the proximity as well as in the IGM region (though the uncertainty can be larger in the wings of the \lya emission line). However, its impact will be slightly diluted at least for the analysis of the ratio of median values of these observed pixel optical depth. \par Ideally, it would be better to have mock spectra for these quasars with a known continuum and then fitting the continuum to those mock spectra to get the statistical and systematical continuum uncertainty.  In the absence of mock spectra for our data-set, we have adopted a conservative estimate of about 10\% for statistical continuum uncertainty (per pixel), based on the detailed analysis carried out in JCS19 using mock spectra for SDSS spectra (with known continuum). This is a conservative estimate for the high quality spectra used here with SNR $\sim$20 and R$\sim$5100.  However, for the correction of systematic uncertainty due to the systematic under-estimation of continuum, we adopted a typical estimate given by  \citet[][]{Faucher2008ApJ...681..831F} as $\Delta C/C_{true} = 1.58\times 10^{-5} (1+z)^{5.63}$ where $\Delta C$ is the difference between the estimated continuum ($C_{est}$) and actual continuum ($C_{true}$). The analysis was performed using mock spectra with spectral properties similar to that of the data used here. A similar continuum correction is also given by JCS19 for SDSS data and by \citet[][]{Becker2007ApJ...662...72B} for HIRES data. We present our results using the continuum fits given by \citet[][]{Lopez_2016refId0} (e.g. see the red-dashed line in Fig.~\ref{Fig:spectra}) after applying the aforementioned correction for systematic shift in the continuum (e.g. see the red-solid line in Fig.~\ref{Fig:spectra}). However, for the sake of completeness, we also discuss our results without applying  this continuum shift in Sect.~\ref{s:Discussion}.

         \begin{figure*}	
          \centering
          \includegraphics[height=10cm,width=15.0cm]{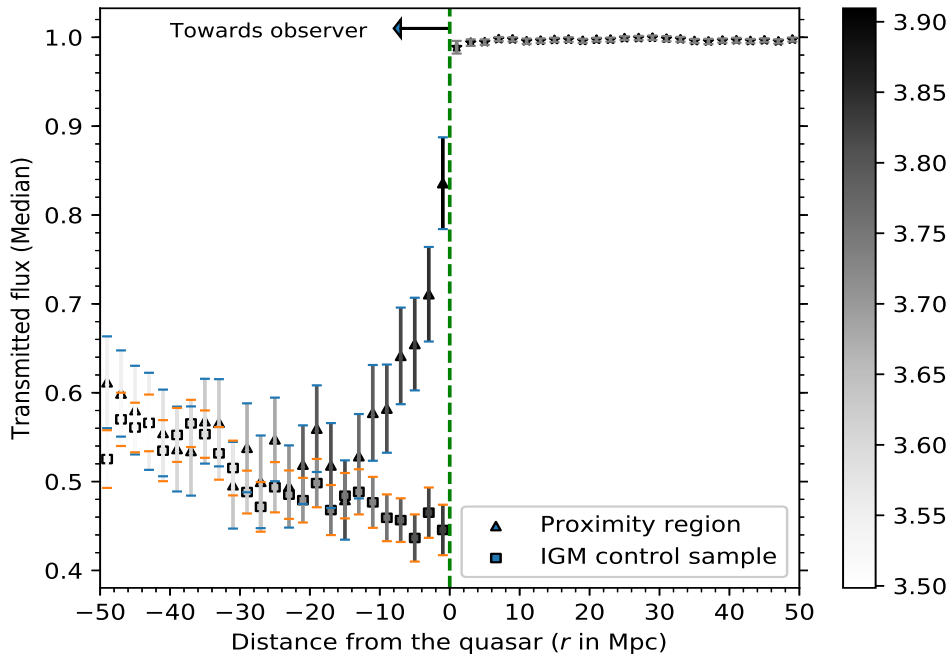}
        \caption {The plot shows the median value of the
          transmitted flux in various 2 Mpc radial distance bins
          from the quasars for the proximity region (triangle) and
          the corresponding control sample (square). The negative
          distances in the x-axis are just sign convention used in
          Eq.~\ref{eq:r_para} to represent the absorbing clouds
          present in the vicinity of the main quasars towards the
          observers. The gray-shades represents the median value
          of the absorption redshifts of pixels used in each
          radial bin. The transmitted flux beyond \lya emission
          line is shown with an asterisk sign (without gray-shades
          according to redshift) to distinguish them from the 
           absorption in the \lya forest. The error bars on the
          median transmitted flux of the proximity region consists
          of redshift uncertainty, flux and continuum measurement
          errors, sightline-to-sightline variance and r.m.s
          statistical errors within each radial distance bin
          (e.g., see Eq.~\ref{eq:f_er_total} and/or
          Sect.~\ref{s:Uncertainties}). The error-bars in the
          control sample are small due to the large sample size (being
          $\sim$ 5 IGM sightlines per main quasar sightline). The
          green vertical dashed line shows the location of the
          main quasars.}
        \label{Fig:rad_flux}
         \end{figure*}

 \subsection{Estimation of quasar's emission redshift}
 \label{s:zq}

 An accurate estimation of the emission redshifts for the quasars is a very crucial aspect for the analysis of the proximity effect. For our entire sample, \citet[][]{Lopez_2016refId0} have given the $z_{em}$, using their robust method based on PCA. Alternatively, as pointed out by many previous studies \citep[e.g., see][and references therein]{Shen2016ApJ...831....7S} that [O~{\sc iii}] emission line is also a reliable estimator of $z_{em}$, which is covered in spectra of $\sim$ 45 sources in our sample. We limit the redshift estimation only to 23 quasars (out of 45 quasars) which has strong enough [O~{\sc iii}] emission lines (equivalent width, EW $>$ 0.3  \AA). Therefore, as a quality check, we fit the doublet of the [O~{\sc iii}] lines ($\sim$ 4959, 5007 \AA) for these 23 sources with double Gaussian to fit both broad and narrow components. Furthermore, to make the fit physical, we have tied the values of redshifts and FWHMs of the doublet lines to have the same values and the intensity ratio to be as 1:3 (i.e., $I_{5007} = 3 \times I_{4959}$) as expected theoretically. We also applied the systemic redshift offset of 48 \kms~as given by \citet[][]{Shen2016ApJ...831....7S} where they assume that Ca~{\sc ii} emission line provides the most reliable systemic redshift. We found that our emission redshift estimate is consistent within a typical dispersion of $\sim$ 150 \kms~with negligible systemic offset with the emission redshifts given by \citet[][]{Lopez_2016refId0} estimated using PCA. 
    
\par Additionally, we also used Mg~{\sc ii} emission line, available for all sources to estimate the redshifts, though the line strength was not high enough (i.e., EW $>$ 0.3) for 23 sources (out of 100). After using double Gaussian fit for these 77 sources and applying a systemic redshift offset of 57 \kms~as given by \citet[][]{Shen2016ApJ...831....7S}, we again found a negligible systemic difference while comparing with PCA based redshifts estimates by \citet[][]{Lopez_2016refId0}. However, we do find a dispersion of $\sim$ 600 \kms~between redshift estimated using Mg~{\sc ii} emission line and PCA based method \citep[see  also,][]{Paris2017A&A...597A..79P}. This suggests that the $z_{em}$ estimated by \citet[][]{Lopez_2016refId0} based on the PCA method is in general consistent with [O~{\sc iii}] and Mg~{\sc ii} emission lines. Finally, for the sake of homogeneity, we have used the $z_{em}$ value as given by \citet[][]{Lopez_2016refId0} for the entire sample. However, to be on the conservative side, we have included a dispersion of 600 \kms~as a typical uncertainty in the individual measured emission redshift value in our further analysis. 
\par Additionally, we calculate the statistical uncertainty in the emission redshift using the Monte-Carlo simulations. For this, at    each pixel with wavelength $\lambda_i$, we have generated flux using Gaussian random distribution with a mean value taken as the   observed flux ($F_i$) and width of the distribution as the observed error on $F_i$. We then estimate the redshift for about 1000 simulated realizations of our sample and found a spread of just 5 \kms~which is negligible in comparison to the above redshift uncertainty of 600 \kms.

  \subsection{ Transmitted flux uncertainties}
  \label{s:Uncertainties}
For estimating the uncertainties in the transmitted flux measurements, we follow a similar procedure as detailed in JCS19 with a brief description below. The first contribution of error in the normalized spectrum is propagated from the flux measurement errors and continuum fitting error ($\Delta F^{fc}_t[\lambda_i]$) at each pixel. As mentioned in Sect.~\ref{s:conti}, we have assumed a  maximum possible uncertainty of 10\% in each pixel as continuum uncertainty as a conservative estimate. It may be noted that there might be a correlation in the continuum placement error within a spectrum. Therefore, the flux error is averaged over the number of pixels in a radial bin w.r.t. the quasars, however, continuum placement error is averaged over the number of spectra contributing in that radial distance bin. This leads to an average of $\Delta F^{fc}_t$ value in each radial distance bin. The second contribution to the error in the normalized flux comes from the r.m.s. scatter of the normalized flux ($\Delta F^{rms}_t$) within each 2 Mpc radial distance bins. 
  
\par Additionally, we include the uncertainty in the median $F_t$ measured in the proximity region due to the typical emission redshift uncertainty ($\Delta F^{z}_t$) of $\sim$ 600 \kms~along our sightlines as discussed in Sect.~\ref{s:zq}. To calculate this, we add a random (Gaussian generator with $\sigma = 600$ \kms) velocity offset to each quasar emission redshift within $\pm$ 600 \kms~range. The distribution of the median transmitted flux computed for 100 such realizations in a radial bin of 2 Mpc is found to be Gaussian with a width of $\Delta F^{z}_t$. Additionally, we also added the sightline-to-sightline variance of the proximity sample ($\Delta F^{var}_t$). Here, we used the bootstrap technique \citep[e.g., see][]{Efron1993} by the random exclusion of sightlines at the cost of repeating other sightlines. We use 100 such realizations to estimate the width of the Gaussian distribution of the median transmitted flux, $\Delta F^{var}_t$, within each 2 Mpc radial distance bins. \par Therefore, the total error on the measured transmitted flux in the proximity region ($\Delta F^{p}_t$) is given by the quadratic sum of all the above-mentioned errors leading to $\Delta F^{p}_t$ for a given 2 Mpc radial distance bin, as,
 \begin{equation}
 \Delta F^{p}_t = \sqrt{(\Delta F^{fc}_t)^2+(\Delta
  F^{rms}_t)^2+(\Delta F^{z}_t)^2+(\Delta F^{var}_t)^2}.
 \label{eq:f_er_total}
 \end{equation}
Here, $\Delta F^{fc}_t$ is of the order of $\sim 0.05$, $\Delta F^{rms}_t$ contributes $\sim 0.006$, $\Delta F^{z}_t$ leads to approximately $\sim
0.001$. Lastly, the $\Delta F^{var}_t$ is of the order of $\sim 0.03$
in the \lya forest. Therefore, the
total error is found to be of the order of $\sim 0.05$ in the
proximity region within each 2 Mpc radial distance bins.
 
\par We follow similar error analysis for the estimation of errors for the control sample after excluding the contribution from redshift uncertainty and sightline-to-sightline variance. Nonetheless, the transmitted flux values of the control sample might be correlated because of the overlap in the comparison samples for quasars at similar emission redshifts. We mitigate such correlations by exclusion bootstrapping, viz., by randomly retaining only one sightline out of the total 5 sightlines used as a control sample for each sightline of the proximity region, and  generated 100 such realizations. The distribution of the transmitted flux in each radial bin is found to be Gaussian. The width of this Gaussian distribution is then added to the error budget of IGM. Moreover, the sample size of the control sample is typically five times larger than the proximity sample (due to our choice of 5 control samples for each main sample) which results in much smaller statistical errors.

\subsection{Transmitted flux statistics}
\label{s:Analysis1}

We have plotted in Fig.~\ref{Fig:rad_flux} median values of $F_t$ within various 2 Mpc radial distance bins, towards and away from the quasars. The negative distances in the x-axis are just sign convention used in Eq.~\ref{eq:r_para} to represent the absorbing clouds present in the vicinity of the main quasars towards the observer. The gray-shades show the median redshift of the pixels contributing to each of the radial distance bin. \par A first noticeable point from Fig.~\ref{Fig:rad_flux} is that the $F_t$ measured at large radial distances from the quasar (i.e., $\geq  14$ Mpc) is consistent with that from the IGM obtained using the control sample (as expected). For each radial distance bin, we have combined all the pixels based on their distances from the quasar falling within that radial bin, but these pixels have absorption redshift spread over a large range. This can result in a gradual slope in the transmitted flux with radial distance for the \lya absorbers as the median absorption redshift of the pixels in a given radial distance bin decreases towards the observer (e.g., see redshift based gray-shade in Fig.~\ref{Fig:rad_flux}). However, we reiterate that our selection of the control samples (see, e.g., Sect.~\ref{s:control_sample}) with exact matching in absorption redshift takes care of such effects due to the similar redshift evolution expected. This is also evident from the similarity of redshift depicted by gray-shade for both the IGM and proximity region. \par 

From Fig.~\ref{Fig:rad_flux}, it can be seen that the proximity effect is significant up to a distance of 12 Mpc with a clear increase in the transmitted flux ($F_t$) as we go closer to the quasar as expected in the classical proximity effect. Also, in the region redward of the quasars  $F_t \sim 1$, which is expected as it represents the continuum region. The consistency of $F_t \sim 1$ can be noticed even in the first radial bin redward of quasars. This indicates that the possible uncertainty in the redshift estimation \citep[as given by][]{Lopez_2016refId0} due to any peculiar motion (if any) are not significant at scales $\geq$ 2 Mpc.

   \begin{figure}	
     \centering
     \includegraphics[height=9cm,width=9.0cm,trim=0 0 0 0]{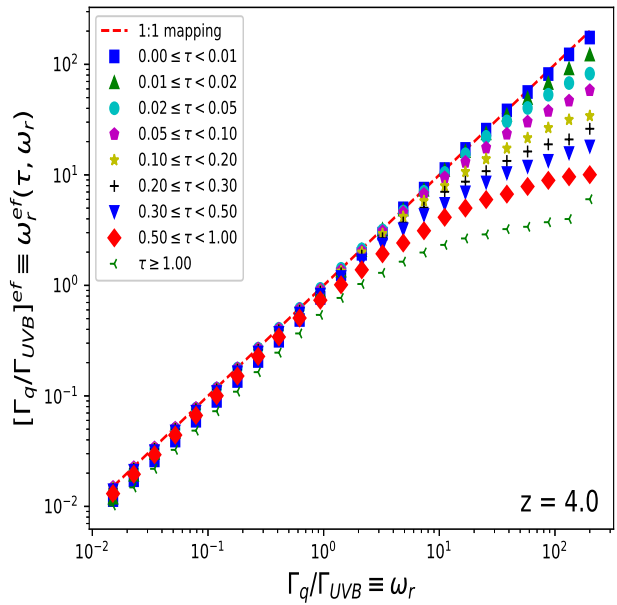}
     \caption {The plot shows the departure of the effective
      ionization correction parameter,
      $[\Gamma_q/\Gamma_{\textsc{\sc uvb}}]^{ef} (\equiv
      \omega^{ef}_r(\tau,\omega_r)$), from its theoretical value of
      $\Gamma_q/\Gamma_{\textsc{\sc uvb}} (\equiv \omega_r$) in
      various pixel optical depth bins (listed in inset) obtained for spectra generated with a resolution mimicking the X-SHOOTER observations.
      }
      \label{Fig:simu}
    \end{figure}

   \begin{figure*}	
  \centering
   \includegraphics[height=7cm,width=8.0cm]{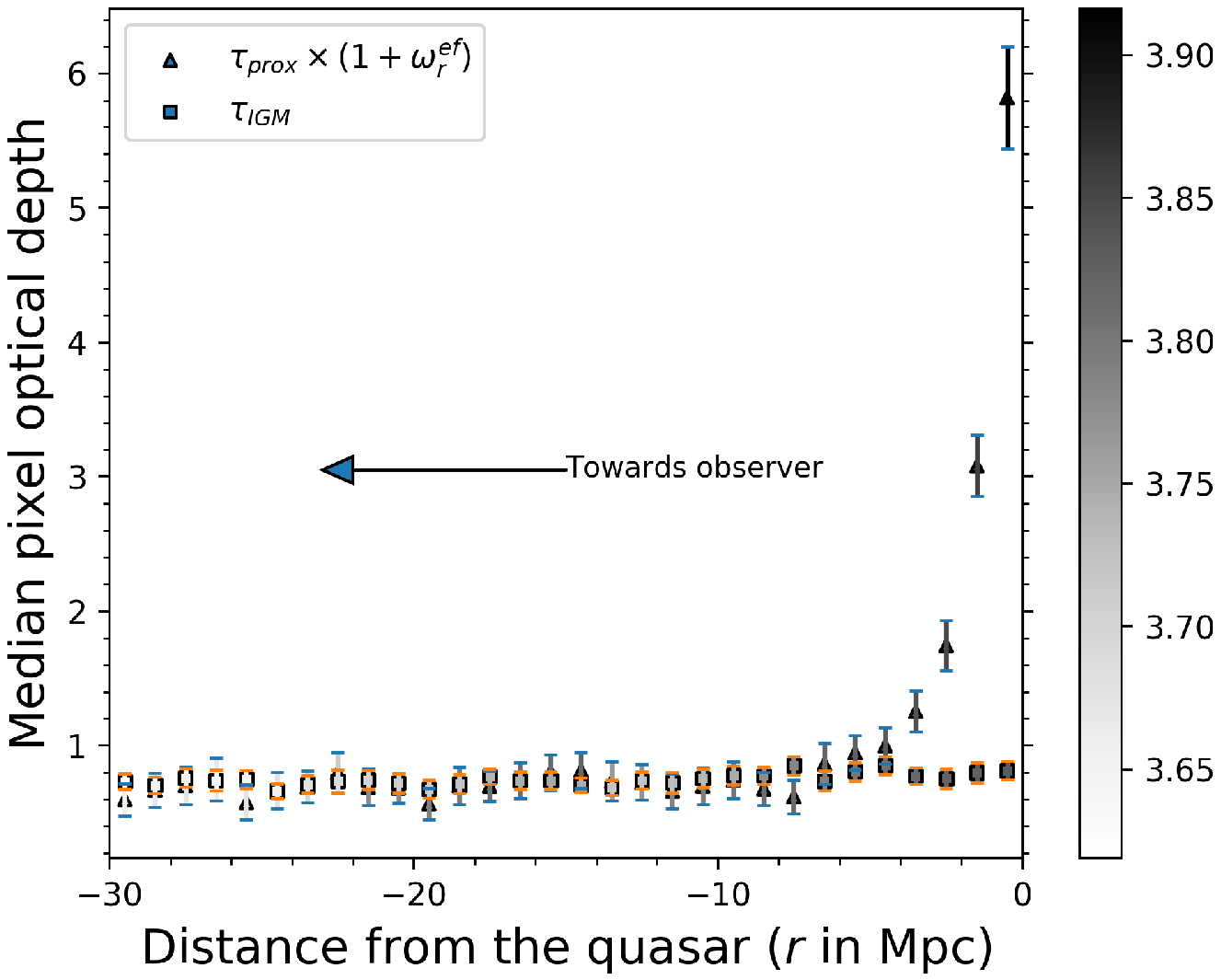}
 \includegraphics[height=7cm,width=8.0cm]{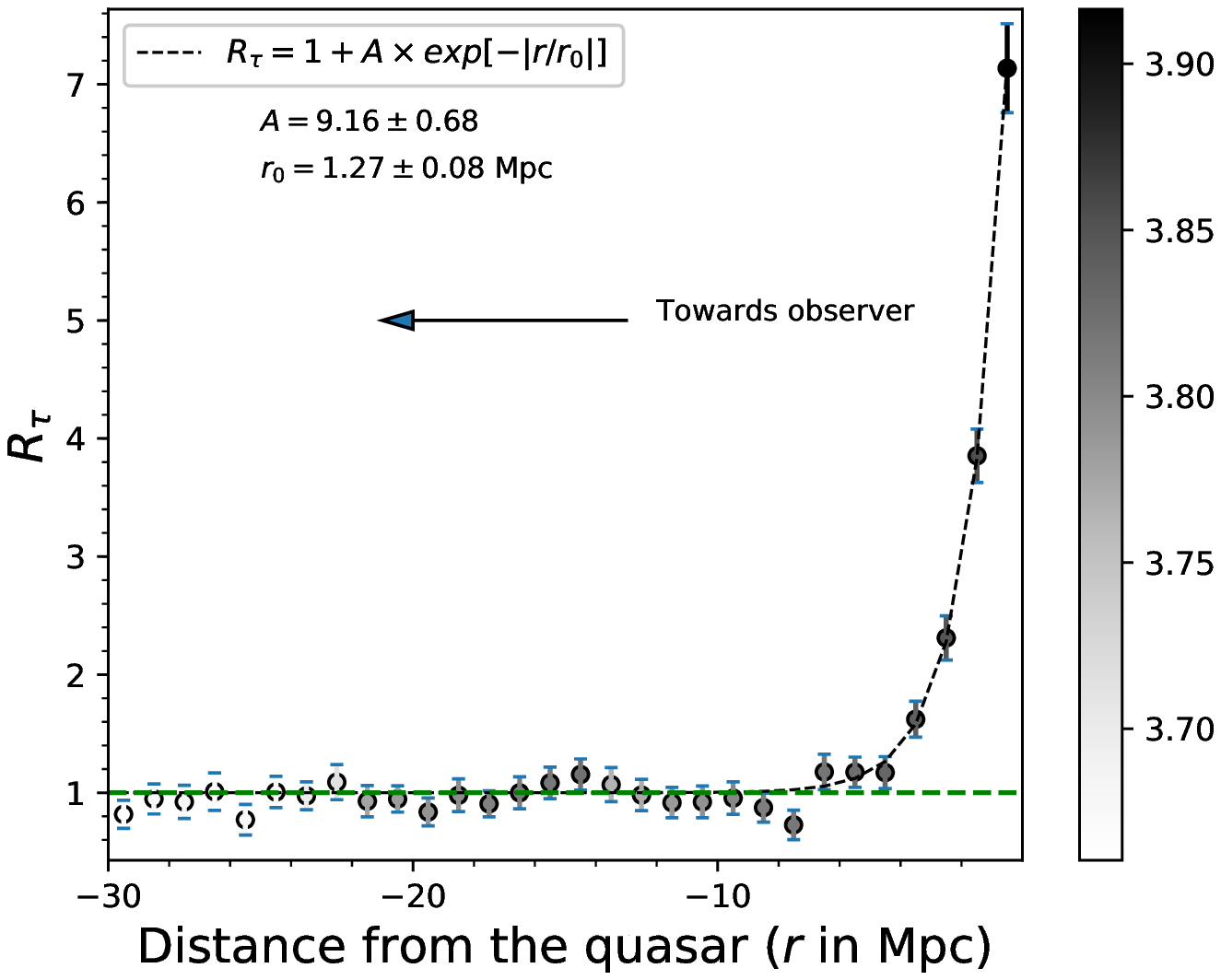}
  \caption {{\it Left panel:} The plot shows the median pixel
      optical depth corrected for quasar's ionization in the proximity
      region (triangle) and the median pixel optical depth of the
      general IGM (square) at different radial distance bin of 1 Mpc.
    The gray-shades represents the median value of the absorption
    redshifts (along the proximity sightline) in each radial distance
    bin. The resultant error bars consist of the flux error from
    photon counting statistics, redshift uncertainty,
    sightline-to-sightline variance and r.m.s statistical error within
    the 1 \Mpc~radial distance bin as also used in
    Fig.~\ref{Fig:rad_flux} along with the error propagation in
    $[1+\omega^{ef}_r(\tau,\omega_r)]$ based on the uncertainty in the
    pixel optical depth.  {\it Right panel:} The plot shows the
    ratio of median  pixel optical depth shown as two curves in
    the left panel (defined as $R_\tau (r)$ in
    Eq.~\ref{eqn:shiftprox2} and~\ref{eqn:ratio}), in a radial distance
    bin of 1 Mpc.}
  \label{Fig:density}
   \end{figure*}

 \subsection{Median pixel optical depth profile around the quasar}
 \label{s:halo_mass}
 \subsubsection{Degeneracy between ionization and excess optical depth at moderate resolution}
 \label{s:ionize}
To lift the degeneracy between the effect of excess ionization from the quasars and the presence of excess pixel optical depth ratio around the quasar in comparison to the IGM, we follow the procedure as detailed in JCS19, with a brief discussion given below. Here, we make use of the UVB measurements from \citet[][]{Khaire2015MNRAS.451L..30K,Khaire2019MNRAS.484.4174K} based on an independent method using an updated comoving specific emissivities of galaxies and quasars at different frequencies (from UV to FIR) and redshifts.  The combined effect leads to shift the observed  pixel optical depth in the proximity region $\tau_{p}$ (e.g., see JCS19 and references therein) as would have been expected in the IGM ($\tau_{\textsc{igm}}$),
\begin{equation}
  \tau_{\textsc{igm}} =
  \tau_{p}\,\frac{[1+\omega^{ef}_r(\tau_p,\omega_r)]}{\left(\rho(r)/\bar{\rho}_{\textsc{igm}}\right)^{\beta}}
  \label{eqn:shiftprox}
\end{equation}
leading to
\begin{equation}
  \Big[\frac{\rho(r)}{\bar{\rho}_{\textsc{igm}}}\Big]^{\beta} = \frac{\tau_{p} \times [1+\omega^{ef}_r(\tau_p,\omega_r)]}{\tau_{\textsc{igm}}} 
  \label{eqn:shiftprox2}
\end{equation}
with $\beta \equiv 2-0.7(\gamma-1)$, with $\gamma$ representing the slope of the temperature (T)-density ($\Delta$) relation viz., T=T$_o\Delta^{\gamma-1}$. Here, $\tau_{p}$ (or $\tau_{p}[\lambda_i]$) is the measured pixel optical depth in the presence of the quasar and $\tau_{\textsc{igm}}$ is the expected pixel optical depth from the IGM which we have estimated using the absorption redshift matched control sample. The term $\tau_{p}\times [1+\omega^{ef}_r(\tau,\omega_r)]/\tau_{\textsc{igm}}$ represents the ratio of the  pixel optical depth corrected for quasar's ionization in the proximity region to that of the  pixel optical depth of the general IGM. We have estimated this ratio within a radial bin of 1 Mpc, by taking median value, i.e.,
\begin{equation}
   \frac{median(\tau_{p}\times  [1+\omega^{ef}_r(\tau,\omega_r)])}{median(\tau_{\textsc{igm}})} \equiv R_\tau (r).
  \label{eqn:ratio}
\end{equation}
  This will be the manifestation of the excess overdensity around the quasars (e.g., see Eq.~\ref{eqn:shiftprox2}) in the radial bin where $R_\tau (r)$ is more than unity, though the mapping between $R_\tau (r)$ and overdensity might require proper calibration using detailed numerical simulations. \par  The $[1+\omega^{ef}_r]$ is equivalent of [$1+\omega_r$] ionization scaling of $\tau_{p}$ after taking into account the effect of moderate spectral resolution (R $\sim$ 5100) and pixel optical depth value where [$1+\omega_r$] is defined as,
\begin{equation}
  \frac{\Gamma_{\textsc{\sc uvb}}(z_{a})+\Gamma_{q}(r,z_{a})}{\Gamma_{\textsc{\sc uvb}}(z_{a})}
  \equiv 1+\omega_r
  \label{eqn:omega}
\end{equation}
with $\Gamma_{\textsc{\sc uvb}}(z_{a})$ \citep[taken
  from][]{Khaire2015MNRAS.451L..30K,Khaire2019MNRAS.484.4174K} and
$\Gamma_{q}(r,z_{a})$ being the H~{\sc i} photoionization rates at the
absorption redshift $z_{a}$ contributed by the UVB and quasar
respectively.  \par In Fig.~\ref{Fig:simu}, we show the departure of
the effective ionization correction parameter,
$[\Gamma_q/\Gamma_{\textsc{\sc uvb}}]^{ef} (\equiv
\omega^{ef}_r(\tau,\omega_r)$), from its theoretical value of
$\Gamma_q/\Gamma_{\textsc{\sc uvb}} (\equiv \omega_r$) in various
pixel optical depth bins at spectral resolution of $\sim$ 5100 (the
typical resolution of our X-SHOOTER spectra). Here, the $\omega_r$ is
used to vary the level of ionization in the simulated IGM spectra
\citep[][]{Gaikwad2018MNRAS.474.2233G}, with pixel optical depth
$\tau_{true}$ as $\tau_{true}/[1+\omega_r]$. The transmitted flux
corresponding to $\tau_{true}$ and $\tau_{true}/[1+\omega_r]$ are
convolved with Gaussian kernel corresponding to the X-SHOOTER
resolution to obtain transmitted flux leading to simulated pixel
  optical depths $\tau_{\textsc{igm}}$ and $\tau_{p}$
respectively. The best fit is computed such that KS-test probability
is maximum for the distribution of
$\tau_{p}\times[1+\omega^{ef}_r(\tau,\omega_r)]$ and distribution of
$\tau_{\textsc{igm}}$ to be similar at each optical depth bin of
$\tau_{p}$, as detailed in JCS19.  In the hydro-dynamical simulation
of IGM used here from \citet[][]{Gaikwad2018MNRAS.474.2233G}, they
have used GADGET-3
simulation{\footnote{http://wwwmpa.mpa-garching.mpg.de/gadget/}}, with
box size of 10 h$^{-1}$ cMpc having 2 $\times$ 512$^3$ number of
particles and mass resolution of $\sim$10$^5$ h$^{-1}$ M$_\odot$. The
UV-background used in this simulation was given by
\citet[][]{Haardt2012ApJ...746..125H}. We note that the UVB used in
our analysis
\citep[i.e.,][]{Khaire2015MNRAS.451L..30K,Khaire2019MNRAS.484.4174K}
is different from the UVB model used in the simulations
\citep[][]{Haardt2012ApJ...746..125H}. This replacement of the UVB
model may alter the overall optical depth distribution of simulated
IGM. However it will not have any significant impact on the above
$\omega_r $ and $\omega^{ef}_r$ relation, as we have used here the
simulations only to calibrate the ionization correction at a given
pixel optical depth (using a small bins of optical depth, e.g., see
Fig.~\ref{Fig:simu}), which will not have any significant dependence
on the overall distribution of the optical depth.
\par It is evident from Fig.~\ref{Fig:simu} that there is a significant departure of $\omega^{ef}_r(\tau,\omega_r)$ from $\omega_r$ at-least for higher optical depth bins and higher values of $\omega_r$.


\subsubsection{Pixel optical depth statistics}

In our further analysis, we calculate the pixel optical depth values (integrated value over the pixel width of $\sim $ 20 \kms) from the transmitted flux $F_{t}(\lambda_i)$ as used in the Fig.~\ref{Fig:rad_flux}, the binning is however reduced to 1 Mpc. We plot in the right panel of Fig.~\ref{Fig:density} the ``ratio of median  pixel optical depth (i.e., $R_\tau(r)$)'' in each 1 Mpc radial bin. The errors are then propagated appropriately as discussed in Sect.~\ref{s:Uncertainties} from each pixel to a given radial distance bin of 1 Mpc. The plot shows that the ratio of median  pixel optical depth increases with a decrease in the radial distance $r$ with a significant excess in the range $-6 \leq r \leq 0$ Mpc. In addition to these errors, uncertainty in the $\omega^{ef}_r(\tau,\omega_r)$ due to measurement uncertainty in
 $\tau$ is also taken into account, as,
\begin{equation}
  |\omega^{ef}_r(\tau+\Delta \tau)-\omega^{ef}_r(\tau-\Delta \tau)|/2.
   \label{eq:wef_t}
\end{equation}

In order to quantify the radial extents around the quasars, where the
$R_\tau(r)$ is significantly higher than unity, we parameterize the
$R_\tau(r)$ versus $r$ curve by a fitting function of the form
$R_\tau(r) = 1+A \times exp(-r/r_0)$. The best fit parameters are
determined by using the $\chi^2$ fit, resulting in $A=9.16\pm 0.68$
and $r_0= 1.27\pm 0.08$ Mpc. The best fit profile is shown in the
right panel of Fig~\ref{Fig:density}, which deviates from the unity
with $ \geq 1\sigma$ for region $-6$ Mpc$\leq r \leq$0 (proper), with
$\sigma$ being the median value of error-bar on $R_\tau(r)$.  In this
6 Mpc region, the integrated value of scaled pixel optical depth
(i.e., $\tau_{p}\times [1+\omega^{ef}_r(\tau,\omega_r)]$ by triangles
in left panel of Fig.~\ref{Fig:density}) is found to be higher by a
factor of 2.55$\pm$ 0.17 in comparison to the corresponding integrated
value of the $\tau_{\textsc{igm}}$ curve (shown by squares in left
panel of Fig.~\ref{Fig:density}). Here the estimated uncertainty is
computed based on the proper error propagation (assuming Gaussian in
nature) from the individual error-bar as shown in the left panel of
Fig.~\ref{Fig:density}. We also used the Kolmogorov-Smirnov test
(KS-test) and found that the null probability for the distribution of
$\tau_{p}\times [1+\omega^{ef}_r(\tau,\omega_r)]$ and
$\tau_{\textsc{igm}}$ within the 6 Mpc to belong to same distribution
is ruled out at 100\% confidence level.

\section{Results and discussions}
\label{s:Discussion}
Studies in the past have established that quasars reside in an overdense gaseous environment based on various techniques such as clustering analysis, cross-correlation etc \citep[e.g., see][]{White10.1111/j.1365-2966.2012.21251.x,Trainor2013ApJ...775L...3T}. Here, we have used a novel technique from our recent study in JCS19 to estimate the ratio of median  pixel optical depth (i.e., $R_\tau (r)$) profile surrounding the high-redshift quasars based on the longitudinal proximity effect. The procedure involves the analysis of the proximity effect using the existing measurements of UVB leading to an estimate of the $R_\tau (r)$ profile around the quasar. The noticeable point in our analysis is that it takes care of the redshift evolution of the  pixel optical depth using a control sample matched in absorption redshift along with similarity in S/N. This procedure surpasses the method of scaling the  pixel optical depth to a reference redshift. The latter has a caveat of scaling continuum noise pixels as well as problematic for the saturated pixels. Additionally, our study also takes into account the effect of the spectral resolution and pixel optical depth while correcting for the effect of quasar's ionization. \par The sample used in this study consists of 85 quasars from the XQ-100 survey covering a redshift range from 3.5 to 4.5. These quasars have Lyman continuum luminosity in the range of 1.06$\times 10^{31}$ to 2.24$\times 10^{32}$ erg s$^{-1}$ Hz$^{-1}$ (e.g., see Fig.~\ref{Fig:z_lum}) with spectral resolution of R $\sim 5100$ and S/N$\sim 30$ (e.g., see Sect.~\ref{s:sample}). We find that the presence of the proximity effect up to a proper distance of 12 Mpc (e.g., see Fig.~\ref{Fig:rad_flux}). It can be noted from  Fig.~\ref{Fig:density} right panel, that the ratio of median  pixel optical depth is found to be higher than unity in the region of $-6 \leq r \leq 0$ Mpc (proper) in the vicinity of the quasar. The integrated value of $\tau_{p}\times [1+\omega^{ef}_r(\tau,\omega_r)]$ (shown by triangle in the left panel of Fig~\ref{Fig:density})  in the $-6 \leq r \leq 0$  region, is found to be higher by a factor of  2.55 $\pm$ 0.17 than that of  the corresponding integrated value of $\tau_{\textsc{igm}}$ (shown by square in the left panel of Fig~\ref{Fig:density}).

\par In our analysis, we have assumed a single value of the spectral index ($\alpha$) for all the sources. However, previous studies have reported a range of values for the spectral indexes. For example, \citet[][]{Stevans2014ApJ...794...75S} estimated $\alpha = 0.83 \pm 0.09$ for 1200-2000 \AA~(rest frame) and $\alpha = 1.41 \pm 0.15$ for 500-1000 \AA~for the composite spectrum obtained by Hubble Space Telescope.  Similarly, \citet[][]{Lusso2015MNRAS.449.4204L} analysis estimates far-ultraviolet spectral index $\alpha = 0.61 \pm 0.01$ and at shorter wavelengths $\alpha = 1.70 \pm0.61$. As pointed out in Sect.~\ref{s:Properties} that in our previous work (JCS19), we have shown that the impact of dispersion of the UV-spectral index is negligible as far as $R_\tau (r)$ is concerned. Similar to JCS19, we have re-estimated our ratio of median  pixel optical depth profile using two extreme values of UV-spectral index viz., 1.96 and 0.56. We found similar $R_\tau (r)$ profile as shown in right panel of Fig.~\ref{Fig:density}. In the distance range of $-6 \leq r \leq 0$ Mpc with the integrated value of $\tau_{p}\times [1+\omega^{ef}_r(\tau,\omega_r)]$ if found to be higher than  integrated value of $\tau_{\textsc{igm}}$  by a factor of 2.28$\pm$ 0.16 and 3.42$\pm$ 0.20 for $\alpha = 1.96$ and $\alpha = 0.56$ respectively.

\par We also note that though in the visual check the continuum fit given by \citet[][]{Lopez_2016refId0} seems to be very robust, but at high redshifts the crowding of \lya absorption can lead to no absorption free region in the \lya forest. In this scenario, the best fitted continuum may be underestimated. In order to take into account the effect of such continuum underestimation, we also apply a systematic continuum shift based on the analysis of \citet[][e.g., see, Sect.~\ref{s:conti}]{Faucher2008ApJ...681..831F}. However, we also repeated the analysis of median transmitted flux (i.e., Fig.~\ref{Fig:rad_flux}) and median  pixel optical depth (i.e., Fig.~\ref{Fig:density}) without applying such continuum shift. We found that the proximity effect is still evident upto a distance of 12 Mpc and the ratio of median  pixel optical depth is significantly higher than unity in the $-5 \leq r  \leq 0$ Mpc.
   
\par We also note that in our recent work JCS19, we have estimated the $R_\tau (r)$ in the longitudinal direction using 181 quasar   pairs from SDSS (relatively smaller spectral resolution, R $\sim 2000$ and S/N $\sim$ 10) sample using the same technique. The longitudinal proximity effect and the $R_\tau (r)$ in that study were found significant up to 4 Mpc. The extent and the significance  level of the detection are comparatively much higher in the present study perhaps due to the higher quality quasar spectra used (R  $\sim$ 5100 and S/N $\sim 30$) and/or probably due to higher redshift and luminosity of the quasars in the sample used here.

\par Such proximity effect is also reported by many of the previous studies. For example, \citet[][]{Guimaraes2007MNRAS.377..657G} also used quasars with a similar redshift range (i.e., $z_{em}>4$) and found evidence for the proximity effect to a distance of $r \sim 15-20$ Mpc. However, in their analysis, while correcting for quasar's ionization they have not considered its dependence on the spectral resolution and pixel optical depth as we have done in our analysis (e.g., see Sect.~\ref{s:halo_mass}).  Similarly, \citet[][]{Calverley2011MNRAS.412.2543C} found proximity effect evident from 3 Mpc to 10 Mpc at the redshift range from 4.5 to 6.5. Here, a tentative increase evident in the extent of the proximity region with an increase in the average emission redshift of the sample could be due to bias in selecting higher luminosity quasars at higher redshifts \citep[see, also][]{Eilers2017ApJ...840...24E}. \par Similarly, in the context of our result that $R_\tau (r) > 1$ for $-6 \leq r \leq 0$ Mpc, which could be the manifestation of the excess overdensity around quasar, we note that, \citet[][]{Scott2000ApJS..130...67S} found the excess overdensity within 1.5 $h^{-1} $ Mpc at 5.5$\sigma$ level. \citet[][]{Rollinde2005MNRAS.361.1015R} claimed detection of overdensities of about a factor of 2 on scales $\sim$ 5 Mpc for quasars at redshift $z \sim 2$. Similarly, \citet[][]{Guimaraes2007MNRAS.377..657G} reported an excess overdensity of the order of 2 on the scales of $\sim$ 10 Mpc. A small minority of quasars in \citet[][]{Aglio2008AA...491..465D} show substantial overdensities of up to a factor of a few, till 3 Mpc. Furthermore, \citet[][]{Dodorico2008MNRAS.389.1727D} found excess overdensity in the region within 4 Mpc from the quasar's position. The main improvement in our present study of the ratio of median  pixel optical depth study is that while applying the correction for the excess ionization of the quasar, we have taken into account the effect of spectral resolution and the pixel optical depth-dependence.
 
\par  Our result that $R_\tau (r) > 1$ is also consistent with the fact that in order to form structures in the universe the collapsed regions should have a mean density higher than its critical value to overcome the Hubble flow \citep[e.g., see,][]{Loeb1995ApJ...448...17L,Rollinde2005MNRAS.361.1015R}. The halo profiles grow via the infall of matter towards their centers and it can extend to a radius much greater than the virial radius \citep[e.g., see,][and references therein]{Faucherlpe22008ApJ...673...39F}. Therefore, the absorbing gas is not at rest with respect to the quasar center resulting in non-zero peculiar velocities. \citet[][]{Faucherlpe22008ApJ...673...39F} showed that this peculiar velocity shifts the absorption redshift of the gas parcel through the Doppler effect and it is much significant compared to the thermal motion of the gas particles. Additionally, \citet[][]{Hui1997ApJ...486..599H} showed from their analysis that these peculiar velocities play an important role in determining the absorption profiles but its effect on the column density distribution is minor. It may be recalled that in our result of excess  pixel optical depth profile as shown in Fig.~\ref{Fig:density}, all possible sources of errors have been considered except the possible impact of the aforementioned peculiar velocity. Using a first-order approximation as explained in \citet[][]{Faucherlpe22008ApJ...673...39F} an in-falling gas parcel will absorb radiation that has \lya frequency (in the quasar rest frame) at proper distance $r' = r - \Delta r$ from the quasar, where $\Delta r = v_\parallel/H(z)$ with $v_\parallel$ representing peculiar velocity ($<0$ for motion in-falling towards the quasar). Assuming the peculiar velocity to be as significant as Hubble flow i.e., $v_\parallel \sim -420$ \kms~at $\left< z_{em} \right>=4$, we obtain $|\Delta r|=1$ Mpc. Therefore, although there could be a dilution of the magnitude of the measured value of $R_\tau (r)$ due to this peculiar velocity, its impact due to $\geq$ 1 Mpc binning in our analysis may not be significant as far as the excess  pixel optical depth ratio profile is concerned. However, in a realistic situation outflow also exists in the vicinity of the quasar, which can lead to the dilution of any such impact of inflow. Furthermore, as we pointed out in Sect.~\ref{s:Analysis1} that the transmitted flux as shown in Fig.~\ref{Fig:rad_flux}, shows a sharp transition without the wing in the transition from the blue-side (i.e., proximity region) to the redward side of the quasars emission line centroid, suggesting a minor effect of peculiar velocity in our analysis. However, it will be important to quantify how the derived density profile will be modified if we include the peculiar velocities using cosmological simulations.

 \begin{figure*}	
  \centering
  \includegraphics[height=7cm,width=8.0cm]{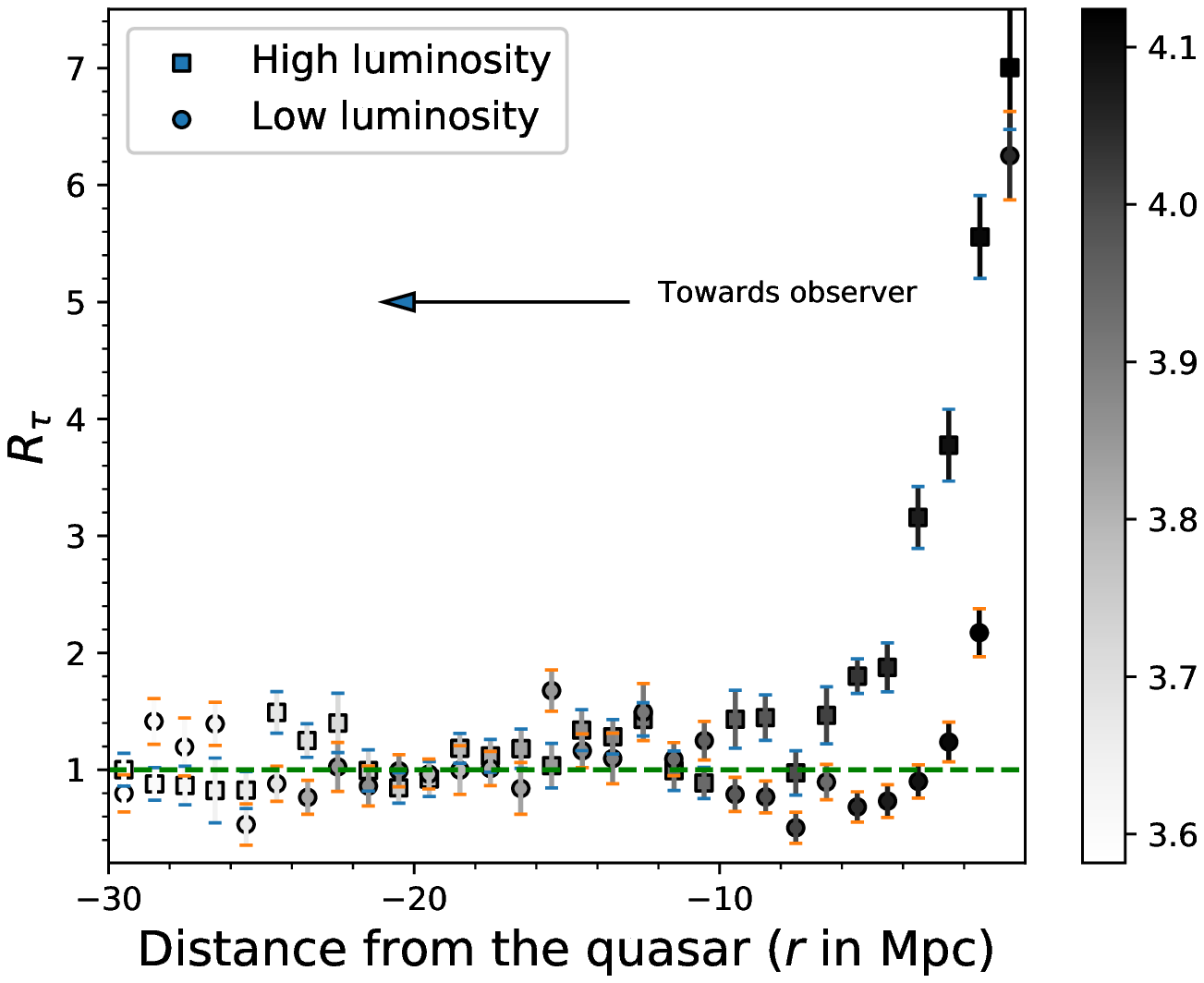}
  \includegraphics[height=7cm,width=8.0cm]{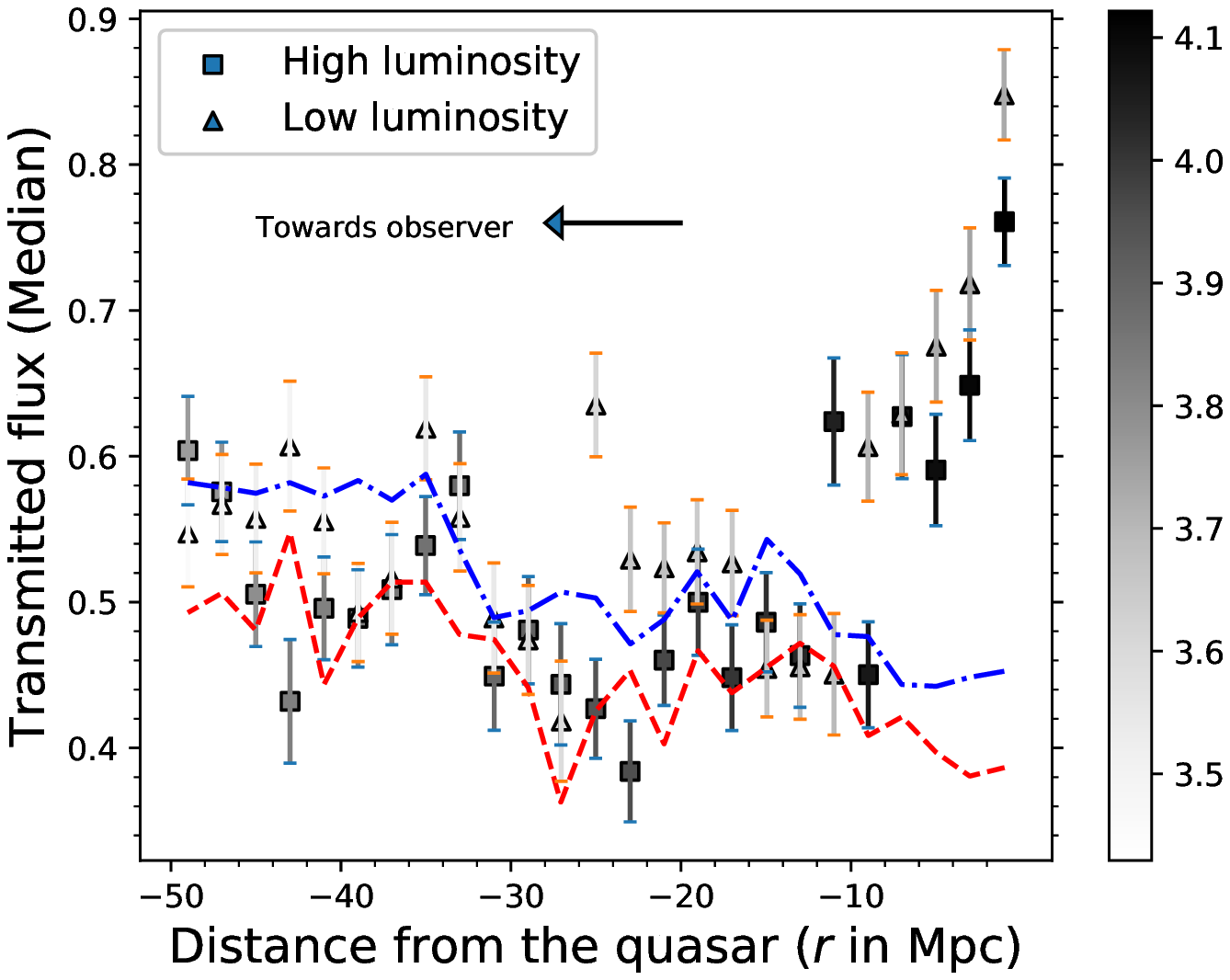}
 \caption {{\it Left panel:} Same as right panel of
   Fig.~\ref{Fig:density}, but for the extreme 25 low and 25 high
   luminosity quasars in our sample (e.g., see text in
   Sect.~\ref{s:Discussion}). The Lyman continuum luminosity was
   calculated using V-band magnitude (e.g., see,
   Sect.~\ref{s:Properties}.)  {\it Right panel:} The plot shows radius
   versus flux (similar to Fig.~\ref{Fig:rad_flux}) for the
   transmitted flux of the subsample of extreme 25 low and 25 high
   luminosity quasars. The transmitted flux of the control sample of
   IGM for the higher-luminosity and the lower-luminosity quasars are
   shown in red-dashed and blue-dot-dashed lines, respectively.}
 \label{Fig:lum}
 \end{figure*}

 \begin{figure*}	
  \centering
    \includegraphics[height=7cm,width=8.0cm]{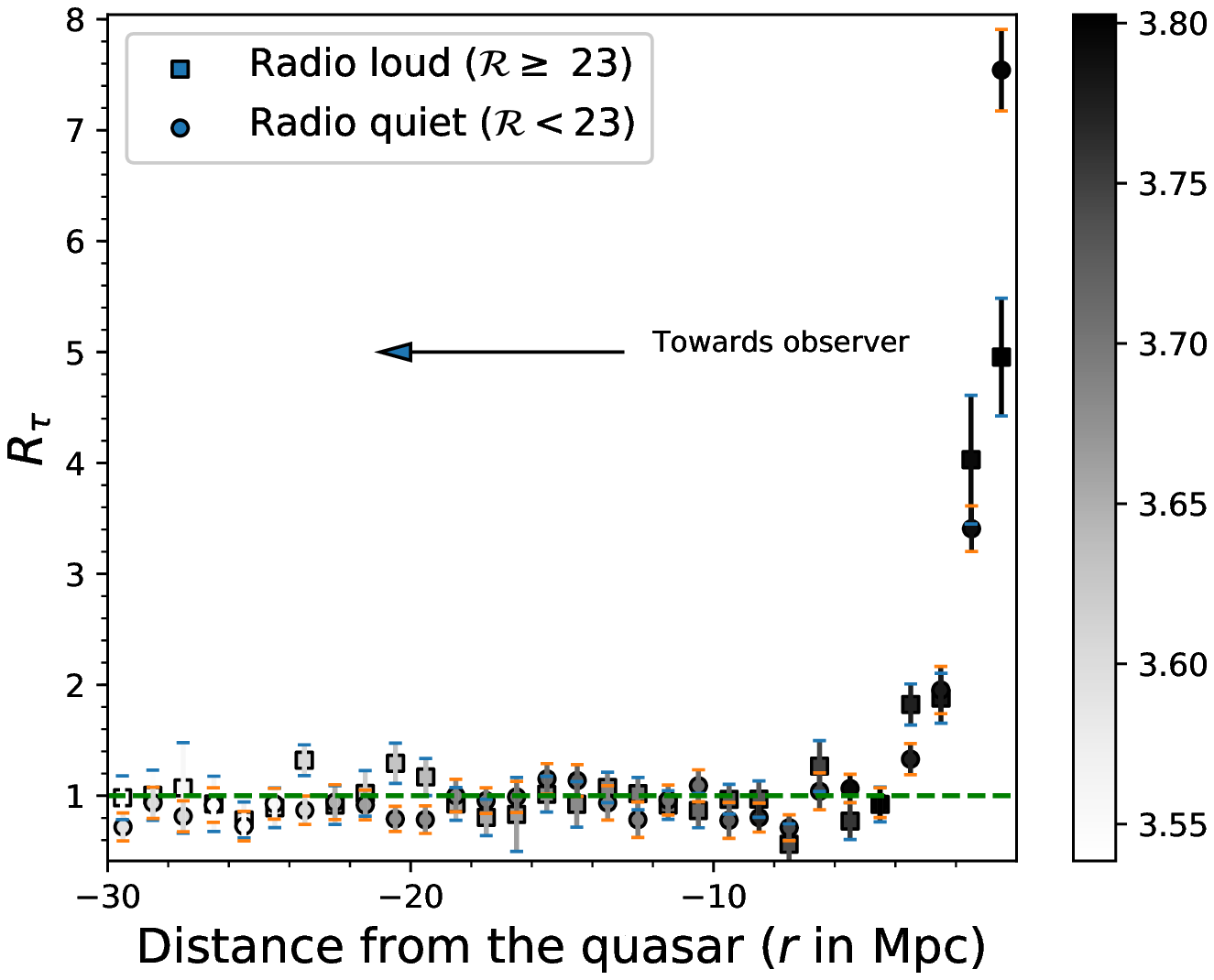}
    \includegraphics[height=7cm,width=8.0cm]{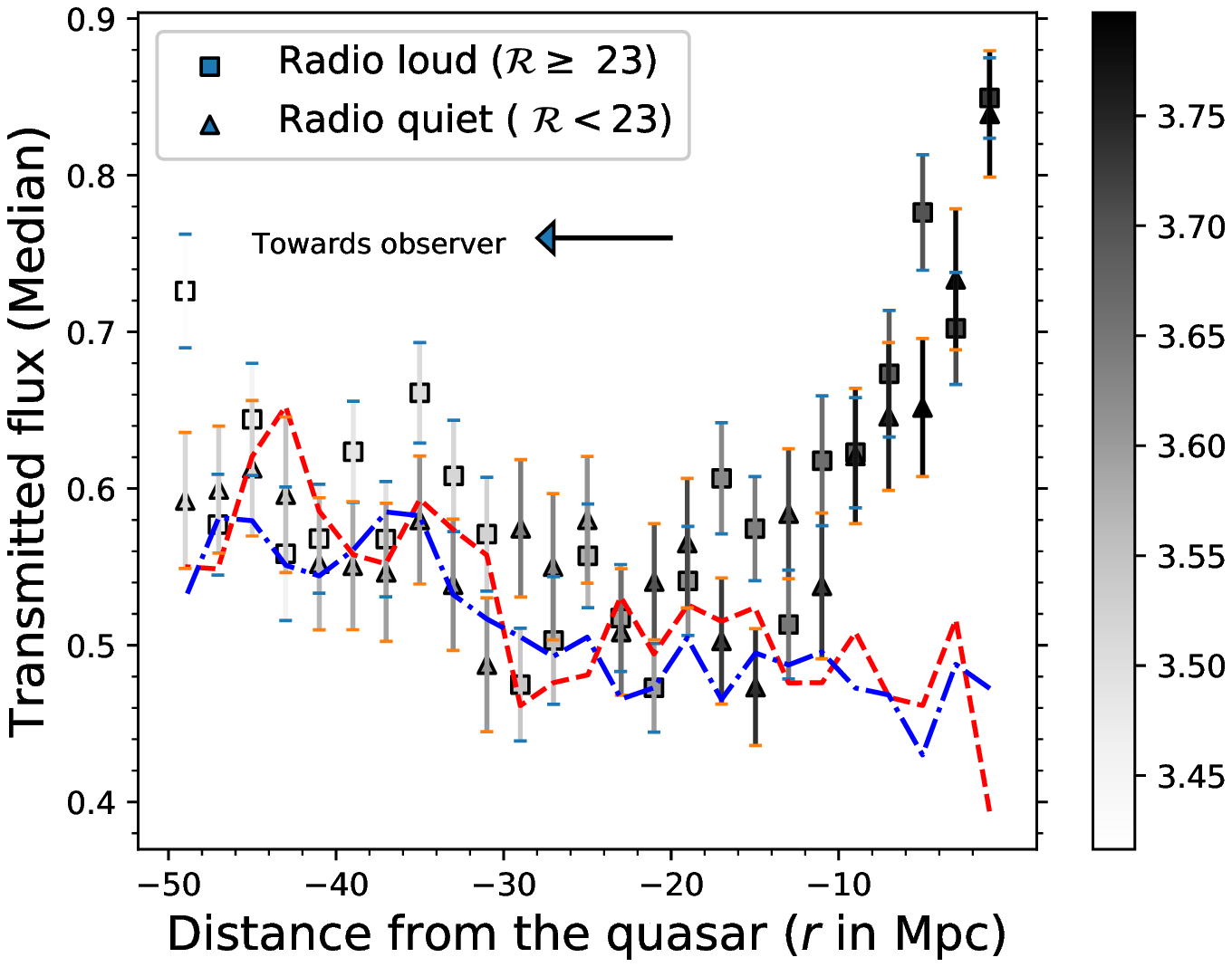}
     \caption{{\it Left panel:} Same as right panel of
       Fig.~\ref{Fig:density}, but for the RL (total 12 quasars with
       $\mathcal{R} \ge$ 23) and RQ (total 50 quasars with
       $\mathcal{R} <$ 23) quasars in our sample. The radio loudness
       parameter ($\mathcal{R}$) was taken from
       \citet[][]{Perrotta2016MNRAS.462.3285P}. {\it Right panel:}
        Same as right panel of Fig.~\ref{Fig:lum},
       but for the RL and RQ subsample of quasars. The transmitted
       flux from the control sample of IGM for RL and RQ quasars are
       shown in red-dashed and blue-dot-dashed lines, respectively.}
\label{Fig:radio}
\end{figure*}

\par Moreover, the $R_\tau (r)$ is also expected to be correlated with the quasar's luminosity, i.e., brighter quasars may live in a region of higher overdensities \citep[e.g. see,][and references therein]{Guimaraes2007MNRAS.377..657G}. In order to check this effect, we devised a subsample of low and high luminosity. In the low and high luminosity bins, we have taken 25 lowest and 25 highest ($\sim$ 29\% of the total sample) luminosity quasars of our sample. As can be seen from Fig.~\ref{Fig:lum} that the excess  pixel optical depth ratio is more pronounced for the subsample of higher luminosity with $\left< logL_{912} \right> = $31.85 ergs s$^{-1}$ Hz$^{-1}$ as compared to the lower luminosity subsample with $\left< logL_{912} \right> = $31.12 ergs s$^{-1}$ Hz$^{-1}$. The integrated value under the $\tau_{p}\times [1+\omega^{ef}_r(\tau,\omega_r)]$ curve (e.g., see Fig.~\ref{Fig:lum}) between  $-6 \leq r \leq 0$ Mpc is found to be higher  by a factor of  3.70 $\pm$ 0.22 and  1.62 $\pm$ 0.15 w.r.t. the corresponding integrated value of $\tau_{\textsc{igm}}$ curve for high and low luminosity subsamples respectively. However, we do notice that higher luminosity bin has a higher median redshift of $\left< z_{em} \right>=4.06$ in comparison to $\left< z_{em} \right>=3.85$ for the lower luminosity bin (e.g., see gray-shade). However, the above $R_\tau (r)$ difference, especially in the $-2 \leq r \leq 0$ Mpc bins, can not be merely explained by the redshift difference in these two subsamples.

\par Additionally, we also used our sample to check for any difference of density profile among radio-loud (RL) and radio-quiet (RQ) quasars \citep[e.g., see][]{Sopp10.1093/mnras/251.1.112,Dunlop10.1046/j.1365-8711.2003.06333.x,Gopal2008ApJ...680L..13G,Retana2017A&A...600A..97R} where the former is generally associated with the presence of jets. Such bi-modality is quantified by a radio-loudness parameter () defined as a ratio between the rest-frame flux densities at 5 GHz and 2500 \AA~\citep[see also,][and references therein]{Ivezi__2002,Kellermann_2016}, with its value R $\ge$ 23 for radio-loud sources as adopted by \citet[][]{Ganguly10.1093/mnras/stt1366}. Out of the 85 quasars in our sample, 62 have information about its radio-loudness parameter with 12 being radio-loud (RL, i.e., $\mathcal{R} \ge$ 23) quasars and 50 being radio-quiet (RQ, i.e., $\mathcal{R} < 23$) quasars \citep[e.g., see][]{Perrotta2016MNRAS.462.3285P}. We plot $R_\tau (r)$ for these two sub-samples (i.e., radio-loud and radio-quiet quasars) as shown in Fig.~\ref{Fig:radio}.  From this figure, it is evident that their $R_\tau (r)$ profile is quite similar in both sub-samples, though the sample size of RL quasars is small here (12 quasars). Here, we also note that the typical luminosity (median) among these two sub-samples is almost similar being 2.23 $\times 10^{31}$ erg s$^{-1}$ Hz$^{-1}$ and 1.55 $\times 10^{31}$ erg s$^{-1}$ Hz$^{-1}$ for the former and latter, respectively.

\par To summarize, our sample of 85 quasars from XQ-100 survey based on X-SHOOTER observations at high R ($\sim$ 5100) and high S/N ($\sim 30$), has allowed us to constrain the  ratio of median  pixel optical depth profile for quasar in redshift range 3.5-4.5 and luminosity range from $1.06\times 10^{31}$ to 2.24$\times 10^{32}$ erg s$^{-1}$ Hz$^{-1}$. Here, while correcting for quasar's ionization, we have properly taken into account the effect of spectral resolution and pixel optical depth using detailed simulations. The ratio of median  pixel optical depth is found to be significant up to 6 Mpc. The area under the curve of  $\tau_{p}\times [1+\omega^{ef}_r(\tau,\omega_r)]$ over 0-6 Mpc range  is found to be higher by a factor of 2.55$\pm$ 0.17  in comparison to the corresponding area under  $\tau_{\textsc{igm}}$ verses $r$ curve. The KS-test  ruled out  the null probability of $\tau_{p}\times   [1+\omega^{ef}_r(\tau,\omega_r)]$ and $\tau_{\textsc{igm}}$ within 6 Mpc to belong to same distribution at 100\% confidence level. This excess factor for  subsample with average luminosity $\left< logL_{912} \right>$ =31.85 (at $\left< z_{em} \right>$=4.06), and $\left< logL_{912} \right>$ =31.12 (at $\left< z_{em} \right>$=3.85),   is found to be  3.70$\pm$ 0.22 and  1.62$\pm$ 0.15 respectively, suggesting its dependence on luminosity.
\par Further improvement to constrain the ratio of median  pixel optical depth profile and its dependence on various parameters such as luminosity, redshift and radio-loudness, will require a much larger sample size from future spectroscopic surveys with high-quality data as is used here from XQ-100 survey.

\section*{Acknowledgements}
The authors thank the anonymous referee for his/her valuable
comments/suggestions which has improved the manuscript. The authors
would also like to thank Dr. Sarah Bosman for the discussion regarding
continuum uncertainties. The authors acknowledge to Dr. Prakash
Gaikwad and Dr. Vikram Khaire for providing the simulated IGM spectra
at $\left< z \right> = 4$ and the latest value of UV-background
radiation, respectively. The authors also like to thank the XQ-100
team for providing the access to the processed data publicly
available, based on observations made with ESO Telescopes at the La
Silla or Paranal Observatories under programme ID(s) 189.A-0424(A),
189.A-0424(B).

\section*{DATA AVAILABILITY}
Based on observations made with ESO
Telescopes at the La Silla or Paranal Observatories under programme ID(s) 189.A-0424(A), 189.A-0424(B). The data used in this article is available at \url{http://archive.eso.org/wdb/wdb/adp/phase3_main/form} as provided by \citet[][]{Lopez_2016refId0}.

\bibliography{references}

\label{lastpage}
\end{document}